\documentclass[pre,twocolumn,showpacs,preprintnumbers,superscriptaddress,floatfix]{revtex4}
\usepackage{graphicx,citesort,psfrag,amsmath,amssymb}
\usepackage[dvips]{color}

\renewcommand{\vec}[1]{\pmb{#1}}
\newcommand{\avg}[1]{\left\langle #1\right\rangle}

\newcommand{\abs}[1]{\left\lvert #1\right\rvert}
\newcommand{\mint}[4]{\int_{#2}^{#3}\!\!#1\,#4}

\newcommand{\Ord}[1]{{\cal O}(#1)}

\newcommand{\ord}[1]{{\text{ord}}(#1)}
\newcommand{\pe}{\perp}
\newcommand{\pa}{\parallel}
\newcommand{\rpe}{r_\perp}

\newcommand{\fopt}{f^{\text{OPT}}}

\newcommand{\fe}{\mathfrak f \,}

\newcommand{\fel}{\vec{f}_{\text{el}}\,}

\newcommand{\fc}{c}
\newcommand{\ludo}{\ell_\text{SWN}}
\newcommand{\lralf}{\ell_\text{EJAM}}
\newcommand{\lpierre}{\ell_\text{BBG}}
\newcommand{\lflower}{\ell_\text{Flower}}
\newcommand{\lp}{\ell_p}
\newcommand{\lpe}{\ell_\perp}
\newcommand{\lpa}{\ell_\|}
\newcommand{\tfin}{t_L^\|}
\newcommand{\tf}{t_\fe}
\renewcommand{\sf}{s_\fe}

\newcommand{\totpe}{t_L^\pe}
\newcommand{\totpa}{t_L^\pa}

\newcommand{\pull}{\text{\emph{Pulling}}}
\newcommand{\epull}{\text{\emph{Towing}}}
\newcommand{\lelo}{\text{\emph{Release}}}

\newcommand{\push}{\text{\emph{Pushing}}}

\preprint{LMU-ASC 60/06}

\begin{document}
\bibliographystyle{apsrev}

\title{Tension dynamics in semiflexible polymers. \\Part~II: Scaling
  solutions and applications }

\author{Oskar Hallatschek} \email{ohallats@fas.harvard.edu}
 \affiliation{Lyman Laboratory of Physics, %
   Harvard University, Cambridge, Massachusetts 02138, USA}
 
 \author{Erwin Frey} \affiliation{Arnold Sommerfeld Center for Theoretical Physics
   and Center for NanoScience, %
   LMU M\"unchen, Theresienstr.~37, 800333 M\"unchen,
   Germany}

\author{Klaus Kroy} 
\affiliation{Institut f\"ur Theoretische Physik,%
  Universit\"at Leipzig, Augustusplatz 10/11, 04109 Leipzig, Germany}

\date{\today}

\begin{abstract}
  In Part~I of this contribution, a systematic coarse-grained
  description of the dynamics of a weakly-bending semiflexible polymer
  was developed. Here, we discuss analytical solutions of the
  established deterministic partial integro-differential equation for
  the spatio-temporal relaxation of the backbone tension. For
  prototypal experimental situations, such as the sudden application
  or release of a strong external pulling force, it is demonstrated
  that the tensile dynamics reflects the self-affine conformational
  fluctuation spectrum in a variety of intermediate asymptotic power
  laws.  Detailed and explicit analytical predictions for the tension
  propagation and relaxation and corresponding results for common
  observables, such as the end-to-end distance, are obtained.
\end{abstract}


\maketitle

\section{Introduction}
\label{sec:intro}
Polymer physics has traditionally focused on very flexible polymers that admit a
highly coarse-grained description in terms of Gaussian chains and exhibit universal
physical behavior that can be explained using methods from statistical mechanics such
as the renormalization group and scaling arguments~\cite{degennes:79}. Further
research has explored new terrain that lies beyond the realm of applicability of this
highly successful approach either because the polymers of interest are too stiff or
because they are subject to extreme forces. Many of these instances have recently
appeared in applications involving biopolymers.  Notorious examples are the nonlinear
mechanical response of DNA~\cite{BustamanteBS03}, which has turned out to be pivotal
to protein-DNA interactions, and the problem of force transduction through the
cytoskeleton~\cite{JanmeyW04,YasuhiroSawada02182002,GaborForgacs06012004}, which is a
major mechanism by which cells explore their environment and react to external
mechanical stimuli.  Clearly, in neither of these situations can theorists contend
themselves with the convenient Gaussian chain representation, but have to resort to
more realistic, yet still schematic descriptions, such as the freely-jointed chain
(e.g.\ for single-stranded DNA) or the wormlike chain model (for double-stranded DNA,
F--actin, microtubules etc.)~\cite{doi-edwards:86,degennes:79}.

Suspicions that this might entail a substantial loss of universality and render
systematic analytical approaches forbiddingly complex have turned out to be
unfounded. The wormlike chain model provides an analytically tractable standard model
for many of the above mentioned new applications, in particular for calculating the
non-equilibrium dynamical response of stiff and semiflexible but weakly bending
polymers to strong external fields.  As established in Part~I of this contribution,
the weakly bending wormlike chain lends itself to a multiple-scale perturbation
theory (MSPT) based on a length scale separation between longitudinal and transverse
dynamic correlation lengths. In the present Part~II, we demonstrate that the
self-affine roughness, acquired by the weakly-bending contour in thermal equilibrium,
plays an analogous role as the more familiar fractal conformational correlations in
the case of flexible polymers~\cite{degennes:79}.  The self-similarity of the static
conformational fluctuations entails self-similar dynamics. It manifests itself in a
variety of \emph{intermediate asymptotic dynamic power laws}.  Apart from the
restriction to polymers with a (locally) rodlike structure, these predictions are as
universal as those of classical polymer physics. They are moreover derived in a
direct way, usually including exact amplitudes, from a controlled perturbation
expansion.

As a major result of the multiple scale theory developed in Part~I, we
obtained a coarse-grained reformulation of the free-draining Langevin
equations of motion of a weakly bending rod in the form of the
deterministic equation
\begin{equation}
  \label{eq:cg-eom}
  \partial_s^2 F(s,t)=-\zeta_\pa \avg{\Delta
    \overline{\varrho}}\left[F(s,\tilde t \leq t),t\right] \;.
\end{equation}
It describes the long wave-length (all time) dynamics of the
time-integrated tension 
\begin{equation}
  \label{eq:integrated-tension}
  F(s,t)\equiv\mint{dt'}{0}{t}f(s,t')
\end{equation} 
with $\zeta_{\parallel}$ being the friction coefficient for longitudinal motion and
$\avg{\Delta \overline{\varrho}}\left[F(s,\tilde t \leq t),t\right]$ the average
release of contour length stored in the transverse undulations up to time $t$.
Written out in terms of the transverse normal mode contributions, the latter reads
\begin{eqnarray}
    \avg{\Delta\overline{\varrho}}(t)&\equiv& \mint{\frac{dq}{\pi\lp}}{0}{\infty}
    \left\{\frac{1}{q^2+f_<}\left(e^{-2q^2[\kappa q^2 t+ F(t)]/\zeta_\perp}-1\right) \right.\nonumber\\
    &&\left.+2 q^2\mint{d\tilde t}{0}{t}e^{-2q^2\left[\kappa q^2(t-\tilde
          t)+F(t)-F(\tilde t)\right]/\zeta_\perp}\right\},
    \label{eq:change-stored-length}
\end{eqnarray}
where $\kappa$, $\lp=\kappa/(k_B T)$ and $\zeta_\perp$ are the bending stiffness,
persistence length and friction coefficient for transverse motion, respectively. The
parameter $f_<\equiv f(t<0)=$const.~allows to take into account a constant
prestress~\footnote{The parameter $\theta$ occurring in the dynamic force extension
  relation of Part~I will be set to one, $\theta=1$, throughout this paper. It
  describes the effect of sudden changes in persistence length, which will be
  discussed elsewhere~\cite{obermayer-kroy-frey-hallatschek3:tbp}.}. Upon inserting
this dynamical force-extension relation into Eq.~(\ref{eq:cg-eom}), we arrived at our
central result, the closed partial integro-differential equation (PIDE) for the
time-integrated tension $F(s,t)$,
\begin{eqnarray}
  \label{eq:pide}
    &\partial_s^2 F(s,t)= \hat \zeta
    \mint{\frac{dq}{\pi\lp}}{0}{\infty}
    \left\{\frac{1}{q^2+f_<}\left(1-e^{-2q^2[ q^2 t+ F(s,t)]}\right)
    \right.& \nonumber \\
    &\left. -2  q^2\mint{d\tilde
        t}{0}{t}e^{-2q^2\left[ q^2(t-\tilde t)+F(s,t)-F(s,\tilde
          t)\right]}\right\}\;.&
\end{eqnarray}
For convenience, we have made the following choice of units: Time and
tension, respectively, are rescaled according to
\begin{eqnarray}
  \label{eq:rescaling-time-tension}
  t&\to&\zeta_\perp t/\kappa \;, \\
  f&\to& \kappa f\;.  
\end{eqnarray}
This corresponds to setting $\kappa\equiv\zeta_\pe\equiv 1$ and $\hat
\zeta\equiv 1/2=\zeta_\pa$. As a consequence all variables represent
powers of length, e.g., $t$ and $f$ are a length$^4$ and a
length$^{-2}$, respectively. 

To leading order in the small contour undulations, the deterministic coarse-grained
tension dynamics, described by Eq.~(\ref{eq:pide}), together with the microscopic
transverse equation of motion, represents a valid reformulation of the constrained
Langevin dynamics of a weakly bending rod subject to a putative pre-stress. The PIDE
Eq.~(\ref{eq:pide}) is the basis not only for discussing the tension dynamics itself,
but also the starting point for analytical and numerical calculations of the
longitudinal and transverse nonlinear response of a weakly bending polymer. It is the
purpose of the present Part~II to treat the former case in detail, while the latter,
somewhat more complex case is reserved for a future
communication~\cite{obermayer-hallatschek2:tbp}. In Sec.~\ref{sec:driving}, we derive
detailed solutions to Eq.~(\ref{eq:pide}) for idealized experimental protocols
involving a representative selection of external fields.  The analytical scaling
solutions obtained for a semi-infinite polymer suddenly pulled (or released) at its
end, reveal the non-trivial short-time phenomenon of tension propagation
(Sec.~\ref{sec:tp}).  At long times, the finite contour length comes in as an
additional characteristic length scale, which gives rise to additional scaling
regimes, discussed in Sec.~\ref{term-rel-tensprop}. To make contact with experiments,
we finally identify the repercussions of the tension dynamics on pertinent
observables like the (projected) end-to-end distance (Sec.~\ref{sec:MSDII}) and
comment on novel experimental perspectives brought up by our analysis
(Sec.~\ref{sec:possible-exps}).

\section{Generic longitudinal driving forces}
\label{sec:driving}
In general, the tension dynamics depends on how the filament is driven externally,
i.e., on the boundary and initial conditions imposed on Eq.~(\ref{eq:pide}).  With
the definition of generic experimental force protocols, this section shall provide a
framework for the analysis of Eq.~(\ref{eq:pide}). We introduce the scenarios
{\pull}, {\epull}, and {\lelo} and report on existing investigations. Apart from
being directly relevant for experiments, we have chosen to consider these scenarios
because of two properties that render them attractive from a theoretical perspective.
Firstly, they correspond to \emph{sudden} changes of the environment and thus do not
introduce an additional delay time scale.  Secondly, since external forces are
assumed to act at the ends, these scenarios only change boundary conditions and leave
the equations of motion unchanged.  In more complicated scenarios, that involve
forces applied not only at the ends, these problems show up as subproblems.  For
instance, if a single point force is applied somewhere within the bulk of a polymer,
the filament can be partitioned into two sections that perceive the external force
only at their ends.

\subsubsection{{\pull}}
\label{sec:pull-def}
The polymer is supposed to be free for negative times, such that it is equilibrated
under zero tension at time zero, i.e., we require
\begin{equation}
  \label{eq:pulling-ic}
  f_<=f(s,t<0)=0 \;.
\end{equation}
Then, for positive times the polymer is pulled in longitudinal direction at both ends
with a constant force $\fe$~\footnote{Throughout, we denote \emph{external} forces or
  force fields by fraktur letters.}. The corresponding external force density
$\fe\left[\delta(s-L)-\delta(s)\right]$ provides the boundary conditions for the
tension (see, e.g., the longitudinal equation of motion in Part~I)
\begin{equation}
  \label{eq:pulling-bc}
  f(s=0,t>0)=\fe \;,\;  f(L,t>0)=\fe \;.
\end{equation}

{\pull} was first considered by Seifert et al.~\cite{seifert-wintz-nelson:96} (SWN).
They predict that a ``large'' tension spreads within a time $t$ a characteristic
length $\lpa(t)=\ludo(t)\equiv \lp^{1/2} (\fe t)^{1/4}$ from the ends into the bulk
of the filament. Their analysis neglects bending forces and thermal forces for the
dynamics, albeit the self-affine thermal initial conditions are used
(\emph{taut-string approximation}).  The contribution of Everaers et
al.~\cite{everaers-Maggs:99} (EJAM) shed light on the linear response to longitudinal
forces. Their simulations established a typical propagation length of
$\lpa(t)=\lralf(t)\equiv \lp^{1/2} t^{1/8}$ for weak forces, which was previously
predicted by Morse~\cite{morse:98II}, and made it plausible by scaling arguments.
Brochard--Wyart et al.~\cite{brochard-buguin-de_gennes:99} (BBG) proposed a theory
for tension propagation claimed to be valid on scales much larger than $\lp$
supposing, however, the weakly bending approximation.  A \emph{quasi-static
  approximation} underlies their analysis, in which the polymer is at any instant of
time considered to be equilibrated with the local tension. Applying their results to
the situation considered here, tension should propagate a distance
$\lpa(t)\propto\lpierre(t)\equiv \lp^{1/2} \fe^{3/4} t^{1/2}$.

Naturally, the scaling arguments used to predict the three different scaling regimes
did not address the crossover and the range of validity. Below, we show that in fact
only two scaling regimes exist.

\subsubsection{{\epull}}
\label{sec:epull-def}

Pulling of a filament can also be studied for time dependent external forces. A
dynamic force protocol of particular experimental relevance is given by the constant
velocity ensemble: The polymer is pulled by a time-dependent external force $\fe(t)$
at the left end such that this end moves with a \emph{constant velocity} $v$
({\epull}).  Besides the growth law of the boundary layer, we wish to understand the
time dependence of the external force.  We will see that both quantities are
proportional to each other because the external force essentially has to drag a
polymer section of length $\lpa(t)$ through the viscous solvent with the constant
velocity $v$, hence $\fe(t)\simeq \hat \zeta v \lpa(t)$.  By measuring the
time-dependent force in a constant velocity experiment one can thus directly monitor
$\lpa(t)$. Possible experimental realizations are outlined in
Sec.~\ref{sec:possible-exps}.

The external force density field corresponding to {\epull} is given by $-\fe(t)
\delta(s)$ with an external force $\fe(t)$ determined to fulfill the requirement that
$\partial_t r_\pa(0,t)=v$. Recall from the equations of motion derived in Part~I,
that, up to terms of order $\Ord{\epsilon}$, the gradient of the tension is given by
the longitudinal friction (as in a rigid rod).  This implies the boundary condition
\begin{eqnarray}
  \label{eq:efpulling-bc}
  \partial_s f(s=0,t>0)&=&-\hat\zeta v+\Ord{\epsilon} \\
  f(L,t>0)&=&0\;. \nonumber
\end{eqnarray}

\subsubsection{{\lelo}}
\label{sec:lelo-def}

{\lelo} refers to the process ``inverse'' to {\pull}: the filament is supposed to be
equilibrated at $t=0$ under a constant pulling force,
\begin{equation}
  \label{eq:let-loose-ic}
  f_<=f(s,t<0)=\fe>0 \;.
\end{equation}
Then, at $t=0$, the external force is suddenly switched off and the filament begins to
relax. The ends are considered to be free for $t>0$,
\begin{equation}
  \label{eq:let-loose-bc}
  f(s=0,t>0)=0 \;,\; f(L,t>0)=0 \;.
\end{equation}
{\lelo} has been discussed by Brochard et al.~\cite{brochard-buguin-de_gennes:99}. According to that work
the characteristic size of the boundary layer, where the tension is appreciably
decreased from $\fe$, should be given by $\lpa(t)\propto\lpierre(t)$ (the same as for
{\pull}). 

Furthermore, Brochard et al.~predict that the tension is relaxed as soon as the
tension has spread over the whole filament yielding a relaxation time $\totpa$ for
the tension that satisfies $\lpa(\totpa)=L$. This is in conflict with what we will
find Sec.~\ref{term-rel-tensprop}, where we identify a novel scaling regime of
homogeneous tension relaxation.

\section{Tension propagation}
\label{sec:tp}
To unravel the physical implications of Eq.~(\ref{eq:pide}) for the scenarios
introduced above, we begin with the tension propagation regime $\lpa\ll L$, where the
total length $L$ of the polymer is irrelevant.  In this regime, it is legitimate to
discuss the dynamics on a (formally) semi-infinite arc length interval $[ 0,
\infty[$.  Problems like {\pull} and {\lelo} still depend on four independent length
scales ($\lp, \fe^{-1/2},s , t^{1/4}$).  Yet, it is shown in
Sec.~\ref{sec:scaling-forms} that, Eq.~(\ref{eq:pide}) is \emph{solved exactly} by a
tension profile that obeys a crossover scaling form depending on only two arguments,
which can be identified as a reduced time and arc length variable, respectively.  In
Sec.~\ref{sec:asymptotic-growth}, we then argue that for asymptotically short ($<$)
and long ($>$) times the scaling function reduces to a function of only one scaling
variable $\xi^{\gtrless}=s/\lpa^{\gtrless}(t)$. Our major results concerning tension
propagation, particularly our classification of tension propagation laws
$\lpa^{\gtrless}(t)$, are summarized in Sec.~\ref{sec:tensprop-summary}.

\subsection{Scaling forms}
\label{sec:scaling-forms}
For each of the generic problems introduced in Sec.~\ref{sec:driving}, we shall see
that the tension profile obeys certain crossover scaling forms, that cannot be
inferred from dimensional analysis. These scaling forms greatly simplify the further
analysis of the tension dynamics by reducing the number of independent parameters.

To solve the equation of motion, Eq.~(\ref{eq:pide}), for a given force protocol, we
proceed in the following way.  A scaling ansatz is postulated and shown to eliminate
the parameter dependence in Eq.~(\ref{eq:pide}) and the boundary conditions after a
suitable choice of length, time and force scales.  These crossover scales turn out to
separate two different regimes, a short- and long-time regime,
respectively~\footnote{Such a crossover was also found in Part~I for the stored
  length under a spatially constant tension from ordinary perturbation theory.}.

Although being ultimately interested in the tension profile $f(s,t)$, it is
convenient to first discuss the \emph{time integrated} tension $F(s,t)$, defined in
Eq.~(\ref{eq:integrated-tension}), because the equation of motion for the tension,
Eq.~(\ref{eq:pide}), is naturally formulated in terms of $F(s,t)$.  The physically
more intuitive quantity $f(s,t) =\partial_t F(s,t)$ is extracted afterwards by a
differentiation with respect to time.
 
We make the following ansatz for the time integral $F(s,t)$ of the tension
\begin{equation}
  \label{eq:sansatz}
  F(s,t)=\fe \tf \; \phi\left(\frac s {\sf } , \frac t {\tf} \right) \;,
\end{equation}
in terms of as yet unknown crossover time and length scales $\tf$ and $\sf$ to be
determined below. While a force scale $\fe$ is given explicitely in the case of
{\pull} and {\lelo} by the pulling/pre-stretching force, a natural force scale
\begin{equation}
  \label{eq:ext-force-towing}
  \fe\equiv \hat \zeta v
  \sf \qquad \text{(\epull)}\:
\end{equation}
for {\epull} is provided not unless $\sf$ is fixed. The combination of variables in
Eq.~(\ref{eq:ext-force-towing}) represents the force necessary to drag a polymer
section of length $\sf$ (longitudinally) through the fluid with the imposed towing
velocity $v$.

As long as the polymer is in equilibrium, $t<0$, the dimensionless scaling function
$\phi(\sigma,\tau)$ for the integrated tension is zero for both pulling scenarios,
but linearly increasing with time for {\lelo} due to the constant pre-stretching
force,
\begin{equation}
  \label{eq:ic-sansatz}
  \phi(\sigma,\tau<0)\equiv \fc \tau=\left\{{0 \;, \atop \tau \;,}
      \qquad {{\pull}/ {\epull}  \atop \lelo }
    \right.
\end{equation}
The constant $c$ entering the initial condition, Eq.~(\ref{eq:ic-sansatz}), is
given by $c= 0$ for {\pull}/{\epull} and $c= 1$ for {\lelo}, respectively.

Since we expect that the signal of a sudden change at the end of the polymer,
i.e.~at $\sigma=0$, takes time to propagate into the bulk of the polymer, which
corresponds to $\sigma\to\infty$, we look for solutions that have a time-independent
stored length at $\sigma\to\infty$.  According to Eq.~(\ref{eq:cg-eom}), this
corresponds to the boundary condition of a vanishing curvature of the tension profile
at infinity,
\begin{equation}
  \label{eq:bc-infty}
  \partial_\sigma^2\phi(\sigma\to\infty,\tau>0) =0
  \;.
\end{equation}
At the origin, the force, respectively, the gradient of the force are prescribed by
the considered experimental setup,
\begin{subequations}
  \label{eq:bc-origin}
  \begin{eqnarray}
    \label{eq:bc-origin-1}
    \phi(\sigma=0,\tau>0)&=&\left\{{\tau \;, \atop 0 \;,}
      \qquad {{\pull}  \atop \lelo } \right. \\
    \partial_\sigma\phi(\sigma=0,\tau>0) &=&-\tau \;, \qquad {\epull}
    \label{eq:bc-origin-2} 
  \end{eqnarray}
\end{subequations}
Inserting the scaling ansatz, Eq.~(\ref{eq:sansatz}), into Eq.~(\ref{eq:pide}) yields
after the variable substitutions $q\to q\sqrt{\fe}$, $t\to \tau \tf$ and $\tilde t
\to \tilde \tau \tf$
\begin{eqnarray}
  \label{eq:pide-inserted-1}
  &&\lp \frac{\fe^{3/2} \tf}{\hat \zeta \sf ^2} \partial_\sigma^2
  \phi(\sigma,\tau) =\\
  &&\mint{\frac{dq}{2\pi}}{-\infty}{\infty}
  \left\{\frac{1}{q^2+\fc}\left[1-e^{-2q^2\left[q^2\tau+
          \phi(\sigma,\tau)\right] \tf \fe^2}\right] \right. \nonumber\\
  &&\left. -2 q^2 \tf \fe^2\mint{d\tilde
      \tau}{0}{\tau}e^{-2 q^2 \left[ q^2 (\tau-\tilde
        \tau)+\phi(\sigma,\tau)-\tilde \tau \phi(\sigma, \tilde
        \tau)\right] \tf \fe^2} \right\}\;.\qquad \nonumber
\end{eqnarray}
By fixing the scales $\tf$ and $\sf $ appropriately,
\begin{subequations} \label{eq:scales1}
  \begin{eqnarray}
    \tf&=&\fe^{-2} \label{eq:scales1-time} \\
    \sf &=&\hat \zeta^{-1/2} \lp^{1/2} \fe^{-1/4}
    \label{eq:scales1-length} \;,
    \end{eqnarray}
\end{subequations}
we can eliminate the parameter dependence of
Eq.~(\ref{eq:pide-inserted-1}),
\begin{eqnarray}
  \label{eq:pide-nondimensional-1}
  &&\partial_\sigma^2 \phi(\sigma,\tau)
    =\mint{\frac{dq}{2\pi}}{-\infty}{\infty} \left\{\frac{1}{q^2+\fc
        }\left[1-e^{-2q^2\left[q^2 \tau+
            \phi(\sigma,\tau)\right]}\right] \right.\nonumber\\
    &&\qquad \left. -2 q^2 \mint{d\hat \tau}{0}{\tau}e^{-2
        q^2 \left[ q^2 (\tau-\hat \tau)+\phi(\sigma,\tau)-
          \phi(\sigma, \hat \tau)\right] } \right\}\;.
\end{eqnarray}
Note that for {\epull}, the conditions in Eqs.~(\ref{eq:scales1-time},
\ref{eq:scales1-length}) imply the scales
\begin{subequations} \label{eq:scales2}
  \begin{eqnarray}
    \tf&=&\left(v^2 \hat \zeta \lp \right)^{-4/5} 
    \label{eq:scales2-time} \\
    \sf&=&\left(v{\hat \zeta}^{3}\lp^{-2}\right)^{-1/5}
    \label{eq:scales2-length}  \qquad ({\epull})\;,
  \end{eqnarray}
\end{subequations} 
since $\fe$ depends on $\sf$ via Eq.~(\ref{eq:ext-force-towing}).

\subsection{Asymptotic scaling}
\label{sec:asymptotic-growth}
We have thus removed the parameter dependence of the differential equation as well as
the boundary and initial conditions. The remaining task is to solve
Eq.~(\ref{eq:pide-nondimensional-1}) for $\phi$ under the initial/boundary conditions
given by Eqs.~(\ref{eq:ic-sansatz},~\ref{eq:bc-infty},~\ref{eq:bc-origin}), and to
extract the tension
\begin{subequations}
  \label{eq:closure-1}
  \begin{eqnarray}
    f(s,t)&=&\partial_t F(s,t) \label{eq:closure-a-1} \\
    &=&\fe \; \varphi\left(\frac s {\sf },\frac t {\tf}\right)
  \end{eqnarray}
\end{subequations}
in terms of the scaling function
\begin{equation}
  \label{eq:close-scaling-form-1}
  \varphi(\sigma,\tau)\equiv\partial_\tau \phi(\sigma,\tau) \;.
\end{equation}

Although this remaining task of finding a solution is analytically not possible in
generality, we now know, at least, that it should be a function of only two
variables, an effective space and time variable $\sigma=s/\sf$ and $\tf$,
respectively. From the conditions in Eqs.~(\ref{eq:scales1-time},
\ref{eq:scales1-length}), it is seen that the scales $\sf $ and $\tf$ are given by a
combination of the characteristic force scale $\fe$ (a length$^{-2}$ in our units)
and the persistence length, and hence not simply a consequence of dimensional
analysis.  The significance of these nontrivial scales is that they mark a crossover
in the behavior of the tension. For it turns out that, the two-parameter scaling form
$\varphi(\sigma,\tau)$ collapses onto one-parameter scaling form in the limit of
large and small arguments, i.e.,
\begin{equation}
  \label{eq:scaling}
  \varphi(\sigma,\tau)\to \tau^\alpha \hat \varphi
  \left(\frac{\sigma}{\tau^z}\right)
  \;,\qquad\text{for } \tau{\ll \atop \gg}1
\end{equation}
with a positive and monotonous scaling function $\hat \varphi(\xi)$
that is bounded as $\xi\to \{0,\;\infty\}$ and exponents $\alpha$ and
$z$ depending on the problem and the limit, i.e., short or long time
limit. Equation (\ref{eq:scaling}) expresses the asymptotic
self-similarity of the tension profile: by stretching the tension
profile at a given time in the arclength coordinate $\sigma$ one
obtains the tension profile at a later time, a property inherited
from the self-affine conformational fluctuation spectrum of the weakly
bending wormlike chain.

Rewriting the scaling variable in Eq.~(\ref{eq:scaling}) as
$\sigma/\tau^z\equiv s/\lpa(t)$ identifies the tension propagation length
\begin{equation}
  \label{eq:def-lf}
  \lpa(t)=\sf \left( \frac{t}{\tf}  \right)^z \;.
\end{equation}
In Tab.~\ref{tab:lf-growth-laws} the actual growth laws $\lpa(t)$ are
tabulated depending on the problem and the asymptotic limit.

Before deriving these growth laws from an asymptotic analysis of
Eq.~(\ref{eq:pide-nondimensional-1}), let us give a simple argument as to what the
exponent $z$ should be on short times. As usual, we assume the crossover should occur
when the scaling variable $\sigma/\tau^z$ is of order one, hence
\begin{equation}
  \label{eq:crossover-requirement}
  \lpa^\gtrless(\tf)\simeq\sf \;.
\end{equation}
If we further assume, that on very short times the propagation length $\lpa^<(\tau)$
of the tension should actually be independent of the external force, we can
immediately infer
\begin{equation}
  \label{eq:short-time-guess}
  \lpa^<(t)\propto \lp^{1/2}t^{1/8} \;,
\end{equation}
which is the correct short time growth law, as will be shown in
Sec.~\ref{sec:short-time-asymptotics}.

The derivations we present in the following are consistent in the sense that we use
assumptions that are a posteriori legitimized by the solutions. In particular, we
exploit Eq.~(\ref{eq:scaling}) as a scaling ansatz in order to derive asymptotic
differential equations for the tension. Those equations are then shown to be indeed
solved by similarity solutions of the postulated type. In addition, let us assume
that the exponent $\alpha$ in Eq.~(\ref{eq:scaling}) is larger than $-1/2$,
\begin{equation}
  \label{eq:alpha-assumption}
  \alpha>-1/2 \;,
\end{equation}
i.e., the tension should increase (decrease) less rapidly than $\tau^{-1/2}$ for
$\tau\to 0$ ($\tau\to\infty$). The assumption is reasonable for {\pull}, because at
the ends, we have $\varphi(0,\tau)=1$ and therefore $\alpha=0$ in this case. It turns
out that Eq.~(\ref{eq:alpha-assumption}) is correct for all considered problems of
tension propagation except for sudden temperature changes, discussed in
Ref.~\cite{obermayer-kroy-frey-hallatschek3:tbp}. As a consequence, the
approximations that are made in the following do not apply to sudden changes in
persistence length, which is an indication that it is an exceptional problem. In
Sec.~\ref{term-rel-tensprop}, where we consider the scaling regime \emph{succeeding}
tension propagation, we encounter an asymptotic regime of {\lelo} characterized by an
exponent $\alpha=-2/3$ as another important exception of
Eq.~(\ref{eq:alpha-assumption}).

Since the central PIDE Eq.~(\ref{eq:pide-nondimensional-1}) is
expressed in terms of the time integrated tension $\phi(\sigma,\tau)$,
it is useful for the following discussion to reformulate the scaling
assumption, Eq.~(\ref{eq:scaling}), in terms of $\phi$,
\begin{equation}
  \label{eq:scaling-phi}
  \phi(\sigma,\tau)\to \tau^{\alpha+1} \hat \phi
  \left(\frac{\sigma}{\tau^z}\right)
  \;,\qquad\text{for } \tau{\ll \atop \gg}1  \;.
\end{equation}

\subsubsection{Short times ($t \ll \tf$)}
\label{sec:short-time-asymptotics} 
\begin{figure}
  \centerline{\includegraphics[width=\columnwidth]
  {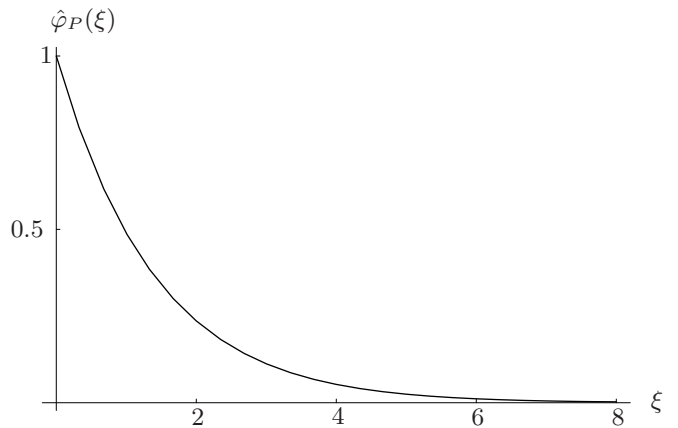}}
  \caption{The scaling function $\hat \varphi_P(\xi)$ describing the shape
    of the tension profile of {\pull} on short times.}\label{fig:small-t-profile}
\end{figure}

In case Eq.~(\ref{eq:alpha-assumption}) holds we can linearize
Eq.~(\ref{eq:pide-nondimensional-1}) for short times in $\phi$. This is at first
sight only correct for small wave numbers that satisfy $q^2\phi=\Ord{ q^2
  \tau^{\alpha+1}}\ll 1$. However, upon a closer inspection of the region of large
wave vectors $q>\tau^{-(\alpha+1)/2}$ that do not allow for a linearization, it is
seen that there is another term in the exponent, $q^4 \tau>\tau^{-2\alpha-1}\gg1$
(for $\tau\ll1$ and $\alpha>-1/2$), which renders the considered exponential
essentially zero.  Therefore, we can approximate Eq.~(\ref{eq:pide-nondimensional-1})
by
\begin{equation}
  \label{eq:pide-linearized}
  \begin{split}
    & \partial_\sigma^2 \phi(\sigma,\tau) \approx \\
    & \mint{\frac{dq}{2\pi}}{-\infty}{\infty} \left\{\frac{1}{q^2+\fc}
      \left[1-\left(1-2q^2\phi(\sigma,\tau)\right)e^{-2q^4
          \tau}\right] \right.\\
    & \left. -2 q^2 \mint{d\hat
        \tau}{0}{\tau}\left[1-2q^2\left(\phi(\sigma,\tau)-\phi(\sigma,\hat
          \tau)\right)\right]e^{-2 q^4 (\tau-\hat \tau)} \right\}\\
    &\qquad\qquad=\mint{\frac{dq}{2\pi}}{-\infty}{\infty}
    \left\{-\frac{\fc}{q^2(q^2+\fc)}\left(1-e^{-2q^4\tau}\right)
    \right. \\
    &\qquad\qquad \left. +2\phi(\sigma,\tau)\left[1-\left(\frac{\fc}
          {q^2+\fc}\right)e^{-2q^4\tau}\right]\right.\\
    &\qquad\qquad\left.- 4q^4 \mint{d\hat \tau}{0}{\tau}\phi(\sigma,\hat \tau) e^{-2
        q^4 (\tau-\hat \tau)} \right\}\;.
\end{split}
\end{equation}
Since $\tau\ll1$ we can neglect the parameter $\fc$ in the denominator of the first
term and set the exponential function in the second term equal to one.  Furthermore,
we observe that the lower bound of the time-integral can be set to $-\infty$ by
defining $\phi(\sigma,\tau<0)\equiv 0$. After the variable substitution $\hat \tau
\to \hat \tau+\tau$ we obtain
\begin{equation}
  \label{eq:pide-linearized-11}
  \begin{split}
    &\partial_\sigma^2 \phi(\sigma,\tau)
    =\mint{\frac{dq}{\pi}}{-\infty}{\infty} \left[\fc \frac{e^{-2q^4
          \tau}-1}{2q^4}+
       \right. \\
    &\left.\frac{q^2 \phi(\sigma,\tau)}{q^2+\fc} - 2q^4\mint{d\hat
        \tau}{-\infty}{0}\phi(\sigma,\hat \tau+\tau) e^{2 q^4 \hat
        \tau} \right] \;.
  \end{split}
\end{equation}
Now we introduce the Laplace transform of $\phi$ according to
\begin{subequations}
  \label{eq:laplace-trafo}
  \begin{align}
    \phi(\sigma,z)& \equiv \mint{d\tau}{0}{\infty}e^{-z \tau}
    \phi(\sigma,\tau) \\
    \phi(\sigma,\tau)&=\mint{\frac{dz}{2\pi i}}{c-i \infty}{c+i
      \infty}e^{z \tau} \phi(\sigma,z)\;,
  \end{align}
\end{subequations}
such that Eq.~(\ref{eq:pide-linearized-11}) in Laplace space reads
\begin{eqnarray}
  \label{eq:pide-linearized-lt-1}
    &&\partial_\sigma^2 \phi(\sigma,z) =
    \nonumber\\
    &&\mint{\frac{dq}{\pi}}{-\infty}{\infty}\left[
       -\frac{\fc}{2q^4}\left(\frac1 z-\frac1{z+2q^4}\right)+
      \frac{q^2 \phi(\sigma,z)}{q^2+\fc}\right.\nonumber\\
    &&\left.-\mint{d\tau}{0}{\infty}e^{-
        z \tau}\mint{d \hat\tau}{-\infty}{0}\phi(\sigma,\hat \tau+
      \tau)(2
      q^4)e^{2q^4 \hat \tau}\right] \nonumber\\
    &&= \mint{\frac{
        dq}{\pi}}{-\infty}{\infty}\left[-\frac{\fc}{z(z+2q^4)}+
     \frac{z \phi(\sigma,z)}{2
          q^4+z}-\frac{\fc \phi(\sigma,z)}{q^2+\fc}\right] \nonumber\\
    &&=2^{-3/4}\left(z^{1/4}\phi-\fc z^{-7/4}\right)-\fc^{1/2}\phi
    \;.
\end{eqnarray}
In the short time limit $z\gg 1$, this reduces to  
\begin{equation}
  \label{eq:linear-pide-laplace}
  \partial_\sigma^2 \phi(\sigma,z)=2^{-3/4}z^{1/4}\phi\;.
\end{equation}
Choosing the decaying solution we find for the boundary conditions in
Eqs.~(\ref{eq:bc-infty},~\ref{eq:bc-origin-1}) of {\pull}
\begin{equation}
  \label{eq:phi-small-t}
  \phi=\phi_P(\sigma,z)\equiv z^{-2} e^{-\sigma (z/8)^{1/8}} \qquad
(\pull) \;.
\end{equation}
Since Eq.~(\ref{eq:pide-linearized-lt-1}) is a linear differential equation, we can
express the solutions for the boundary conditions of {\epull} and {\lelo} in terms of
$\phi_P$,
\begin{subequations}
  \label{eq:letloose-from-pulling}
  \begin{align}
    \phi&=z^{-2}-\phi_P(\sigma,\tau) \qquad (\lelo) \\
    \phi&=\mint{d\hat \sigma}{\sigma}{\infty}\phi_P(\hat
    \sigma,\tau)\qquad (\epull) \;.
  \end{align}
\end{subequations}
Ultimately, we are interested in the tension 
\begin{equation}
  \label{eq:tension-from-phi}
  \varphi_P(\sigma,\tau)=\partial_\tau
  \phi_P(\sigma,\tau) \;,
\end{equation}
which is given by the inverse Laplace transform of $z
\phi_P(\sigma,z)$,
\begin{eqnarray}
  \varphi_P(\sigma,\tau)&=&\hat\varphi_P\left(\frac{\sigma}{\tau^{1/8}}\right)   \label{eq:lt-inverse-a}
  \\
  \hat\varphi_P(\xi)&=&\mint{\frac{dz}{2\pi i z}}{-i \infty+\epsilon}{i\infty+\epsilon}e^{-\xi
      (z/8)^{1/8}+z }\;. \label{eq:lt-inverse-b}
\end{eqnarray}
After deforming the contour of integration such that it encloses the branch cut at the
negative real axis, the integral in Eq.~(\ref{eq:lt-inverse-b}) becomes
\begin{eqnarray}
  \label{eq:small-time-result}
  \hat \varphi_P(\xi)&=&\mint{\frac{dx}{\pi}}{0}{\infty}\frac8x \sin\left[\xi x\sin{\frac \pi 8} 
  \right] e^{-\xi x \cos{\frac \pi 8}}\nonumber\\
  & &\times\left( 1-e^{-8x^8}  \right) \;,
\end{eqnarray}
which is to our knowledge not tabulated, but can be easily evaluated numerically, see
Fig.~\ref{fig:small-t-profile}.  Upon using the known Laplace transform,
\begin{equation}
  \label{eq:power-law-back}
  \frac{\Gamma(\nu)}{z^\nu}=\mint{d\tau}{0}{\infty}e^{-z \tau} \tau^{\nu-1}\;,\quad
    \Re \nu>0\;,
\end{equation}
and Taylor expanding the integrand of Eq.~(\ref{eq:lt-inverse-b}) one obtains an
expansion of $\varphi_P(\xi)$ that is particular useful for small $\xi$,
\begin{equation}
  \label{eq:small-sigma-exp}
   \hat \varphi_P(\xi)=\sum_{{n=0\atop n/8\notin\mathbb{N}}}^{\infty}
   \frac{(-2^{-3/8}\xi)^n}{n!\Gamma(1-n/8)}\;.
\end{equation}
With an absolute error less than one percent the scaling function is approximated by
an exponential,
\begin{equation}
  \label{eq:approx-lr}
  \hat \varphi_P(\xi)\approx \exp\left(-\frac{\xi}{2^{3/8}\Gamma(7/8)}
  \right) \;,
\end{equation}
where the pre-factor of $\xi$ in the exponent is the initial slope
$\partial_\xi \hat\varphi\vert_{\xi=0}$ of the scaling function.

The tension profiles of the problems under consideration are all
related to the scaling function $\hat \varphi_P(\xi)$,
\begin{subequations}
  \label{eq:letloose-from-pulling-2}
  \begin{align}
    f(s,t)&=\fe \hat\varphi_P\left(\frac s {\lpa(t)}\right)\qquad (\pull) \\
    f(s,t)&=\fe \left[1-\hat\varphi_P\left(\frac s
        {\lpa(t)}\right)\right]
    \qquad (\lelo) \\
    f(s,t)&=\hat \zeta v \lpa(t)\mint{d\xi}{s/\lpa(t)}{\infty}
    \hat\varphi_P(\xi)\qquad (\epull) \label{eq:efp-t<1} \;, 
  \end{align}
\end{subequations}
where the boundary layer at time $t$ has the typical size
\begin{equation}
  \label{eq:lambda-short-time}
  \lpa(t)=\hat \zeta^{-1/2} \lp^{1/2} t^{1/8}\propto\lralf(t) \;.
\end{equation}
The scaling was anticipated in Eq.~(\ref{eq:short-time-guess}) and in an ad hoc
scaling argument presented in Part~I. The result for {\epull},
Eq.~(\ref{eq:efp-t<1}), shows that the force at the grafted end, at $\xi=0$, scales
like $\hat \zeta v \lpa(t)$. This scaling will be shown to hold also in the long-time
limit and can be understood in the sense that the graft has to balance only the drag
arising within the boundary layer, since the bulk of the filament is not moving
longitudinally. Thus, measuring the force at the grafted end, we can monitor the
spreading of the tension. This gives special experimental relevance to the {\epull}
scenario.  In particular, we predict
\begin{equation}
  \label{eq:force-at-graft}
  f(0,t)=\hat \zeta v \lpa(t)2^{3/8}/\Gamma(9/8)
\end{equation}
for the force at the grafted end on short times, $\tau\ll1$. The
pre-factor in Eq.~(\ref{eq:force-at-graft}) has been found by
evaluating the remaining integral in Eq.~(\ref{eq:efp-t<1}), which
yields a Taylor series very similar to Eq.~(\ref{eq:small-sigma-exp}),
\begin{equation}
  \label{eq:efp-t<1-full}
  \mint{d\hat\xi}{\xi}{\infty}
    \hat\varphi_P(\hat \xi)=2^{3/8}\sum_{{n=-1\atop 
n/8-1\notin\mathbb{N}}}^{\infty}
   \frac{(-2^{-3/8}\xi)^{n+1}}{(n+1)!\Gamma\left(1-n/8\right)}\;.
\end{equation}

\subsubsection{Long times ($t \gg \tf$)}
\label{sec:long-time-asymptotics}
The present subsection deals with the dynamics of the tension on a semi-infinite
filament on asymptotically long times. We identify and interprete the terms
dominating the stored length release in this limit. Neglecting subdominant terms in
the continuity equation, Eq.~(\ref{eq:cg-eom}), results in differential equations
that can be solved by similarity solutions.

The right hand side of the nondimensionalized PIDE,
Eq.~(\ref{eq:pide-nondimensional-1}), represents the negative change of stored length
in adapted units. It can be written as the sum of two terms, $A$ and $B$, where
\begin{subequations}
  \begin{align}
    \label{eq:A-def}
    A&\equiv\mint{\frac{dq}{2\pi}}{-\infty}{+\infty}\frac 1 {q^2+\fc
    }\left[1-e^{-2q^2\left[q^2 \tau+
          \phi(\sigma,\tau)\right]}\right]\;,\\
    \label{eq:B}
    B&\equiv-\mint{\frac{dq}{2\pi}}{-\infty}{\infty} 
    2 q^2 \mint{d\hat \tau}{0}{\tau}e^{-2 q^2 \left[ q^2
        (\tau-\hat \tau)+\phi(\sigma,\tau)- \phi(\sigma, \hat
        \tau)\right] }\;.
  \end{align}
\end{subequations}
We already pointed out in Part~I that the term $A$ can be interpreted as the
``deterministic relaxation'' of stored length (for the fictitious situation ``T=0'',
i.e.~no thermal noise, for $t>0$). The term $B$ describes the increase in stored
length due to the thermal kicks and is strictly positive. We analyze both terms
separately.

For pre-stretched initial conditions ($\fc=1$) the long-time limit
of $A$ follows from setting the  exponential to zero,
\begin{equation}
  \label{eq:A-let-loose}
  A\stackrel{\tau\gg1}{\to}\mint{\frac{dq}{2\pi}}{-\infty}{+\infty}
  \frac1{q^2+1}=\frac 1 2\;, \quad \text{for }\fc=1 \;.
\end{equation}
The term $A$ for $\tau\to \infty$ is nothing but the initially stored length, which
is completely relaxed after a purely deterministic relaxation. 

The same reasoning cannot be applied in cases of a tension-free
initial state (both pulling problems): setting the exponential to zero
in Eq.~(\ref{eq:A-def}) for $\fc=0$ yields an infrared divergence. The
exponential has to be retained to render the integrand finite at small
wave numbers. For the dominant small wave numbers one can, however,
neglect the term $2 q^4 \tau$ in the exponent because it is small
compared to the term $2q^2\Phi$, so that we arrive at the asymptotic
expression
\begin{equation}
  \label{eq:a-pulling}
  \begin{split}
    A\stackrel{\tau\gg1}{\to}&\mint{\frac{dq}{2\pi}}{-\infty}{+\infty}
    \frac1{q^2}
    \left[1-e^{-2q^2\phi(\sigma,\tau)}\right]\\
    &=\sqrt{\frac2\pi} \sqrt{\phi(\sigma,\tau)}\;,\quad \text{for } \fc=0 \;.
  \end{split}
\end{equation}
I.e. the ``deterministic relaxation'' is dominated by the tension term
and bending can be neglected (as heuristically assumed by Seifert et
al.~\cite{seifert-wintz-nelson:96}). A more formal justification for
Eqs.~(\ref{eq:A-let-loose},~\ref{eq:a-pulling}) is given in
App.~\ref{sec:A}.

The term $B$, describing the stored length generated by the thermal
kicks, takes for asymptotically large $\tau\gg1$ the form
\begin{equation}
  \label{eq:B-large-times}
  B\stackrel{\tau\gg1}{\to}-\mint{\frac{dq}{2\pi}}{-\infty}{\infty}
  \frac 1 {q^2+\partial_\tau \phi(\sigma,\tau)}
  =-\frac1{2\sqrt{\partial_\tau\phi(\sigma,\tau)}}
\end{equation}
independent of the initial conditions. This is shown in App.~\ref{sec:B} upon using
the scaling assumptions in Eqs.~(\ref{eq:scaling-phi},~\ref{eq:alpha-assumption}).
The result, Eq.~(\ref{eq:B-large-times}), should not come as a surprise. It simply
represents the (negative) stored length of a stiff polymer equilibrated at the
(rescaled) tension $\partial_\tau \phi$. For a constant force, i.e., $\partial_\tau
\phi=$const., it is obvious that the stored length should saturate for long times at
the corresponding equilibrium value.  But also if the tension is varying slowly
enough in time ($\alpha>-1/2$) the ``noise-generated'' stored length can be
considered as quasi-statically equilibrated with the tension.

Finally, we combine the ``relaxed stored length'' expressed in $A$ and
the ``noise-generated stored length'' $B$.

\paragraph{Pulling and Towing:} For $\fc=0$ and $\alpha>-1/2$ the term $A\propto
\sqrt{\phi}\propto\tau^{\alpha/2+1/2}$ is much larger than $B\propto
(\partial_\tau\phi)^{-1/2}=\Ord{\tau^{-\alpha/2}}$. I.e. the effects of noise can be
neglected on long times (as presumed by Seifert et
al.~\cite{seifert-wintz-nelson:96}). In this limit, the thermal noise is merely
relevant in preparing the initial state. The relaxation after force application for
$\tau\gg 1$ is purely mechanical, like for a pulled string that is initially prepared
with some contour roughness (\emph{taut-string approximation}). If we replace the
right-hand-side of Eq.~(\ref{eq:pide-nondimensional-1}) by the asymptotic form of
$A$, Eq.~(\ref{eq:a-pulling}), we obtain the partial differential equation
\begin{equation}
  \label{eq:pulling-large-time-pde}
  \partial_\sigma^2\phi(\sigma,\tau)=\sqrt{\frac2\pi} 
  \sqrt{\phi(\sigma,\tau)}
\end{equation}
for the dynamics of the integrated tension $\phi$.  

Eq.~(\ref{eq:pulling-large-time-pde}) represents a Newtonian equation of motion for a
particle moving in a conservative force field $\propto\sqrt{\phi}$ and can be
integrated straightforwardly. In the case of {\pull} the solution, which satisfies
the boundary conditions in Eqs.~(\ref{eq:bc-infty},~\ref{eq:bc-origin-1}) of {\pull} (in
particular $\phi(0,\tau)=\tau$), is given by
\begin{equation}
  \label{eq:pulling-large-time-phi}
  \phi(\sigma,\tau)=\tau\left[\sigma/(72\pi\tau)^{1/4}
        -1\right]^4
\end{equation}
for $\sigma<(72\pi \tau)^{1/4}$ and $\phi=0$ otherwise. The dimensionless tension
$\varphi$ is derived from $\phi$ via differentiation, see
Eq.~(\ref{eq:close-scaling-form-1}), and obeys a scaling form,
\begin{equation}
  \label{eq:pulling-large-time-varphi}  
  f(s,t)=\fe \varphi(s,t)=\fe \hat\varphi\left(\frac s {\lpa(t)}\right)\;,
\end{equation}
with a typical boundary layer size proportional to $\ludo(t)$,
\begin{equation}
  \label{eq:lambda-large-time}
  \lpa(t)=t^{1/4}\hat 
  \zeta^{-1/2}\lp^{1/2}\fe^{1/4} 
\end{equation}
and a scaling function $\hat \varphi$ given by
\begin{equation}
  \label{eq:scalingfct-large-times}
  \hat\varphi(\xi)=\left[1-\xi/(72\pi)^{1/4}\right]^3 
\end{equation}
for $ \xi<(72 \pi)^{1/4}$ and $\hat \varphi=0$ otherwise. {\epull} starts with the same initial conditions ($\fc=0$) but with
different boundary conditions, Eqs.~(\ref{eq:bc-infty},~\ref{eq:bc-origin-2}).
Again, we have to solve Eq.~(\ref{eq:pulling-large-time-pde}), but now
under the boundary condition $\partial_\sigma \phi(0,\tau)=-\tau$. The
solution is
\begin{equation}
  \label{eq:phi-electroforesis}
  \phi(\sigma,\tau)=\tau^{4/3}\left( \frac{9\pi}{32}  \right)^{1/3}\left[
      \frac{\sigma}{(18\pi\tau)^{1/3}}-1  \right]^4  
\end{equation}
for $\sigma<(18\pi\tau)^{1/3}$ and $\phi=0$ otherwise. This implies a tension profile
of
\begin{equation}
  \label{eq:tension-orig-electroforesis}
  f(s,t)=\hat \zeta v \lpa(t)\; \hat\varphi\left(\frac s 
    {\lpa(t)}\right)\;,
\end{equation}
with the typical boundary layer size
\begin{equation}
  \label{eq:lambda-ef-large-time}
  \lpa(t)=    t^{1/3}\lp^{2/3}(v/\hat \zeta)^{1/3}  \;,
\end{equation}
and the scaling function $\hat \varphi$ given by
\begin{equation}
  \label{eq:scalingfct-towing-large-times}
  \hat\varphi(\xi)=\left(\frac{2\pi}{3}
      \right)^{1/3}\left[1-\xi/(18\pi)^{1/3}\right]^3 
\end{equation}
for $ \xi<(18 \pi)^{1/3}$ and $\hat\varphi=0$ otherwise. As on short times, the
absolute value of the reduced tension at the left end is proportional to the size of
the boundary layer. Its precise value is predicted to be
\begin{equation}
  \label{eq:left-end-tension-electroforese}
  f(0,t)=\left(2\pi/3  \right)^{1/3}\hat \zeta v \lpa(t)\propto t^{1/3} \;,
\end{equation}
and should be directly accessible to single molecule experiments.

\begin{figure}
  \includegraphics[width=\columnwidth]{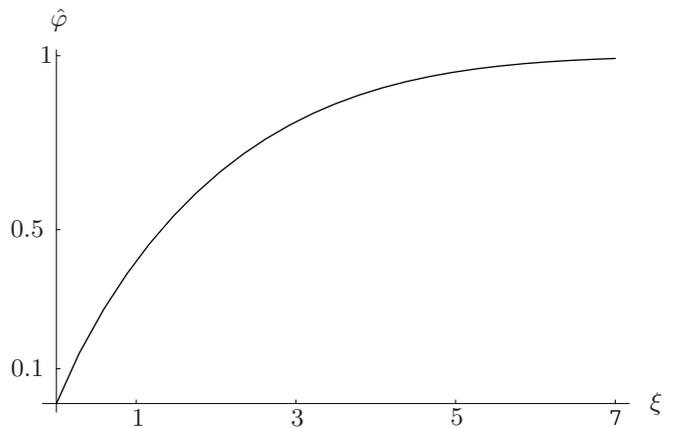}
  \caption{The scaling function $\hat \varphi(\xi)$ for 
    {\lelo} obtained by a numerical solution of
    Eq.~(\ref{eq:scfct-ode}).}\label{fig:degennes-scfct}
\end{figure}

\paragraph{Release:}
The stored length release is asymptotically given by
\begin{equation}
  \label{eq:letting-loose-rho-release-large-t}
  -\avg{\Delta\overline{\varrho}}\propto A+B\stackrel{\tau\gg1}{\sim} \frac 1 2- 
  \frac1{2\sqrt{\partial_\tau \phi(\sigma,\tau)}}\;.
\end{equation}
This expression for the change in stored length can also be directly obtained if one
assumes that the filament was at any time equilibrated with the current tension
$\varphi=\partial_\tau \phi$ (\emph{quasi-static approximation}).  Our derivations
show, that this assumption, used by Brochard et
al.~\cite{brochard-buguin-de_gennes:99}, is only valid in the long time limit for the
scenario of {\lelo}. The dynamics of the integrated tension is then described by
\begin{equation}
  \label{eq:pde-letting-loose-long-t}
  \partial_\sigma^2\phi=\frac 1 2-\frac1{2\sqrt{\partial_\tau 
      \phi(\sigma,\tau)}}\;,
\end{equation}
or, in terms of the tension $\varphi=\partial_\tau \phi(\sigma,\tau)$,
\begin{equation}
  \label{eq:pde-letting-loose-long-t-2}
  \partial_\sigma^2\varphi=\frac1 4 \varphi^{-3/2}\partial_\tau\varphi\;.
\end{equation}
The solution satisfying the correct boundary conditions,
Eqs.~(\ref{eq:bc-infty},~\ref{eq:bc-origin-1}), is given by a scaling form
\begin{equation}
  \label{eq:scform-letting-loose-long-time}
  \varphi(s,t)=\hat \varphi\left(\frac s {\lpa(t)}\right) \;,
\end{equation}
with the typical boundary layer size now growing like
\begin{equation}
  \label{eq:lambda-letloose-large-t}
  \lpa(t)=t^{1/2}\hat\zeta^{-1/2}\lp^{1/2}\fe^{3/4} \propto \lpierre(t)
\end{equation}
and a scaling function $\hat\varphi(\xi)$ satisfying the ordinary differential
equation
\begin{equation}
  \label{eq:scfct-ode}
  \partial_\xi^2\hat\varphi=-\frac1 8 \xi   
  {\hat\varphi}^{-3/2}\partial_\xi\hat\varphi \;.
\end{equation}
The scaling function depicted in Fig.~\ref{fig:degennes-scfct} was already obtained
numerically in Ref.~\cite{brochard-buguin-de_gennes:99}. The slope at the origin is
  $\partial_\xi \hat \varphi\vert_{\xi=0}\approx0.6193$. 

\subsection{Tension propagation (summary)}
\label{sec:tensprop-summary}

This section summarizes the picture of tension propagation that emerges from the
above solutions of the dynamical equation for the tension, Eq.~(\ref{eq:pide}), for
sudden changes in boundary conditions.

For the considered problems {\pull}, {\epull} and {\lelo}, we have shown that the
tension profile in units of $\fe$ has to obey a crossover scaling form
$\varphi(\sigma,\tau)$ depending on a reduced time variable $\tau\equiv t/t_\fe$ and
a reduced arc length variable $\sigma\equiv s/s_\fe$. The scaling function $\varphi$
describes how sudden changes of the tension at the ends spread into the bulk of the
polymer.  Its crossover structure and the expressions for $t_\fe$ and
$s_\fe\approx\lpa(t_\fe)$ are consistent with our heuristic discussion of {\pull} in
Part~I. In the limits $\tau \ll1$ and $\tau\gg1$ the function $\varphi(\sigma,\tau)$
assumes a simple (one-variable) scaling form 
\begin{equation}
  \label{eq:tension-scform}
  \varphi\sim \tau^{\alpha_{\gtrless}}
\hat\varphi^{\gtrless} [\sigma/\tau^{z_{\gtrless}}] \;.
\end{equation}
The notation $\gtrless$
indicates that the asymptotic form of $\hat \varphi$, $\alpha$, and $z$ will
generally not only depend on the kind of external perturbation applied, but will also
differ for times $t\gtrless t_\fe$.  Rewriting $\sigma/\tau^z\equiv s/\lpa$
identifies the tension propagation length $\lpa\equiv s_\fe\,\tau^z$.

For $t\ll t_\fe$ Eq.~(\ref{eq:pide}) could be linearized in $f$ and the scaling
function $\hat \varphi^<$ was obtained analytically. Whereas $\hat \varphi^<$ depends
on the considered force protocol, the corresponding exponent $z_<=1/8$ is independent
of the boundary conditions. As established by our heuristic discussion of {\pull},
this is due to the relaxation of modes with Euler forces $\lpe^{-2}\gg \fe$ much
larger than the external force, for which the equilibrium mode spectrum is hardly
perturbed by the external force. The self-affinity of the equilibrium mode spectrum
translates into a self-similar relaxation dynamics. The dynamic exponent $z$ for the
growth of the boundary layer could already be anticipated from requiring
$\hat\varphi^<$ to become $\fe-$independent as in linear response, see
Eq.~(\ref{eq:short-time-guess}). The short-time dynamics for strong external force is
thus closely related to the linear response. Note, however, that the limit $\fe\to0$
is problematic, as it does not interchange with $\epsilon\to0$.  Our identification
of arc length averages with (local) ensemble averages in Part~I breaks down for
$\fe<(\zeta/\lp)^{1/4}t^{-7/16}$, where fluctuations in the tension become comparable
to its average value.  In fact, extending Eq.~(\ref{eq:pide}) to linear response
amounts to an uncontrolled factorization approximation $\avg{f r_\pe^2}\to
\avg{f}\avg{r_\pe^2}$.  Even in the stiff limit the linear longitudinal dynamic
response remains an open problem. The limit where fluctuations in the tension become
important and its consequences will be detailed in Sec.~\ref{sec:tens-fluct}.

\begin{figure}[t]
  \includegraphics[width=\columnwidth,height=.5\columnwidth
  ]{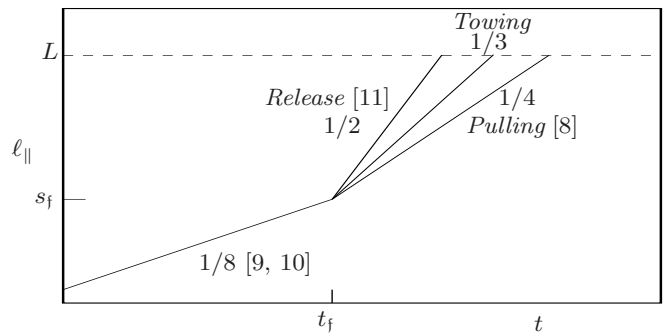}
  \caption{Schematic of the tension propagation laws $\lpa(t)\propto t^z$ on a
    double-logarithmic scale.  At $t_\fe=\fe^{-2}$ they crossover from a universal
    short-time regime to (problem-specific) tension-dominated intermediate
    asymptotics, except for weak forces, $\fe < \lp^2/L^4$. The propagation ends when
    $\lpa(t)\approx L$.}
  \label{fig:lf}
\end{figure}

For $t\gg t_\fe$ the dynamics becomes nonlinear in the external force and starts to
depend on the kind of external perturbation and on how precisely it is applied to the
polymer.  Previously predicted power laws were recovered from Eq.~(\ref{eq:pide}) by
employing different approximations to its right hand side.  In the \emph{taut-string
  approximation} of Ref.~\cite{seifert-wintz-nelson:96}, one neglects for $t>0$
bending and thermal forces against the tension, i.e., one drops the $q^4-$term in the
relaxation time $\tau_q=q^4+f q^2$ of a mode with wave number $q$ and sets
$\lp\to\infty$ for positive times (i.e.~$\theta=0$).  The complementary
\emph{quasi-static approximation} of Ref.~\cite{brochard-buguin-de_gennes:99} amounts
to the omission of memory effects, i.e., to the assumption of instantaneous
equilibration of tension and stored length (as it would be the case for vanishing
\emph{transverse} friction, $\zeta_\pe\to 0$). In cases, where either of these
approximations applies, a power-law dispersion relation combines with a self-affine
mode spectrum to produce self-similar tension dynamics.  Our analysis of
Eq.~(\ref{eq:pide}) showed that either of these approximations becomes rigorous in
the intermediate asymptotic regime defined by $t\gg t_\fe$, $\lpa\ll L$. The
quasi-equilibrium approximation applies to {\lelo} and the taut-string approximation
to {\pull}. We could rule out the applicability of the \emph{taut-string
  approximation} for {\lelo} and of the \emph{quasi-static approximation} for
{\pull}~\cite{brochard-buguin-de_gennes:99} and {\epull}. The ``pure'' scenarios of
self-similar dynamics are summarized in Tab.~\ref{tab:lf-growth-laws} and
Fig.~\ref{fig:lf}.

\begin{table}
  \caption{Asymptotic growth laws for the dynamic 
    size $\lpa(t)$ of the tension boundary
    layer.}
  \label{tab:lf-growth-laws}
  \begin{center}
    \begin{tabular}{c|c|c}
      Problem&$t\ll \tf$&$ \tf\ll t \ll t_L^\pa$\\ \hline
      {\pull}&${\displaystyle \sqrt{ \lp /\hat \zeta }\; 
        t^{1/8}}$&${\displaystyle
        \sqrt{ \lp /\hat \zeta }\; \fe^{1/4}t^{1/4}}$\\
      {\epull}&${\displaystyle \sqrt{\lp/ \hat \zeta  }\; 
        t^{1/8}}$&${\displaystyle
        (\lp^2 v/\hat \zeta)^{1/3}t^{1/3}}$\\
      {\lelo}&${\displaystyle \sqrt{ \lp/ \hat \zeta  }\;t^{1/8}}$
      &${\displaystyle
        \sqrt{ \lp / \hat \zeta  }\;\fe^{3/4}t^{1/2}}$\\
    \end{tabular}
  \end{center}
\end{table}

\section{Terminal stress relaxation} 
\label{term-rel-tensprop}
Up to now, we have considered the growth of the boundary layer in a
stiff polymer that has a (formally) semi-infinite arc length
parameter space, $s\in [0,\infty [$, which is an idealization. However,
the foregoing discussion obviously applies to a polymer of
\emph{finite} length $L$ for sufficiently short times: As long as the
size of the boundary layer is much smaller than the total length $L$
the presence of a second end is irrelevant to the boundary layer
at the first end.  The time where the boundary layers span the whole
polymer marks the crossover to a new behavior. For definiteness, we
define the crossover time $\totpa$ by
\begin{equation}
  \label{tension-rel-time-tp}
  \lpa(\totpa)\equiv L \;.
\end{equation}
What happens for $t>\totpa$? The straightforward way to answer this question is to
solve for the intermediate asymptotics of the PIDE, Eq.~(\ref{eq:pide}), for a polymer
of finite length. The finiteness of $L$ amounts to replacing the boundary condition
$\partial_s^2 f(s\to\infty)=0$ by the correct problem-specific one, i.e., by
\begin{eqnarray}
  \label{eq:bc-tension-rend}
  f(L,t>0)&=&\fe\qquad {\pull} \nonumber\\
  f(L,t>0)&=&0 \qquad {\lelo}, {\epull}.\nonumber
\end{eqnarray}
One could now proceed as in Sec.~\ref{sec:asymptotic-growth} by identifying proper
scaling forms and extracting their asymptotic behavior. As compared to the
semi-infinite polymer limit, this procedure is more complicated for a finite polymer
because of the additional scaling variable $\lpe(t)/L$.  Therefore, we prefer to take
the following short-cut, which consists in two steps.

\subsubsection{Trivial tension profiles for $t\gg t_\star$.}  In the heuristic
analysis of Part~I, we found that ordinary perturbation theory (OPT) should become
valid for times larger than some crossover time $t_\star$. With the exception of
{\epull}, which we discuss separately below, the introduced scenarios treat both ends
equally, such that the tension profile in OPT has a trivial time and arc length
dependence, namely
\begin{equation}
  \label{eq:OPT-tension}
  \fopt(s,t)=f(0,t)=\text{const.}\;,
\end{equation} 
up to small terms of order $\Ord{\epsilon}$. For constant tension we
can easily extract the longitudinal dynamics from
Eq.~(\ref{eq:change-stored-length}). The corresponding
predictions for the evolution of the end-to-end distance will be
discussed in Sec.~\ref{sec:tension-relaxed}.

\subsubsection{Possibility of homogeneous tension relaxation for $\totpa \ll t \ll
  t_\star$.}  Up to now, we have argued that the tension propagates for $t\ll \totpa$
and is constant in time and space for $t\gg t_\star$. There remains the question
whether there is a non-trivial regime of homogeneous tension relaxation in the
time- interval $[\totpa;t_\star]$. Using the systematic approach outlined in
App.~\ref{sec:tstar} to determine $t_\star$, we actually find for most of the
problems that $t_\star\simeq \totpa$, i.e., there is no scaling regime between
tension propagation and the stationary tension profiles dictated by OPT.  The
{\lelo}-scenario, however, provides an important exception, as it allows for a time
scale separation $\totpa\ll t_{\star}$, as we demonstrate explicitely below.  For
intermediate times the tension relaxation is shown to exhibit a novel behavior with
an almost parabolic tension profile and an amplitude that slowly decays in time
according to a power law.

\subsection{{\lelo} for large pre-stretching force}
\label{sec:homogeneous-relaxation}
Let us first determine the time $t_\star$, at which OPT becomes valid, for {\lelo}.
In the OPT regime, the tension should be so small, that we can calculate the change in
stored length accurately by means of Eq.~(\ref{eq:change-stored-length}) (with the
pre-stretched initial conditions of {\lelo}, i.e.~$f_<=\fe$) under the assumption of
a vanishing tension, $\fopt=0$,
\begin{subequations}
  \label{eq:stole-lelo-OPT}
  \begin{eqnarray}
    &&\avg{\Delta \overline{\varrho}}(t)=\mint{\frac{dq}{\pi\lp}}{0}{\infty}
    \left\{\frac{1}{q^2+\fe}\left(e^{-2q^4 t}-1\right)\right.\nonumber\\
    &&\qquad\left.+2
      q^2\mint{d\tilde t}{0}{t}e^{-2q^4 (t-\tilde
        t)}\right\} \label{eq:stole-lelo-OPT-1}\\
    &&=\mint{\frac{dq}{\pi\lp}}{0}{\infty} \left\{\left(
        \frac{1}{q^2}-\frac{1}{q^2+\fe} \right)\left(1-e^{-2q^4
          t}\right) \right\} \label{eq:stole-lelo-OPT-2}\\
    &&=\frac{t^{1/4}}{\lp}\mint{\frac{dq}{\pi}}{0}{\infty}\\
    &&\qquad    \times\left\{\left( \frac{1}{q^2}-\frac{1}{q^2+\sqrt{t/\tf}}
      \right)\left(1-e^{-2q^4} \right)
    \right\}\qquad\label{eq:stole-lelo-OPT-3}  \\
    &&\stackrel{t\gg\tf}{\sim}\frac{2^{3/4}} {\Gamma(1/4)}
    \frac{t^{1/4}}{\lp}\;.\label{eq:stole-lelo-OPT-4}
  \end{eqnarray}
\end{subequations}
Here, we have substituted $q\to q t^{-1/4}$ and replaced $\fe$ by $\tf^{-1/2}$ in
order to obtain Eq.~(\ref{eq:stole-lelo-OPT-3}). In the final step, we took the
long-time limit $t\gg\tf$.  Let us now determine the order of magnitude of the
variation $\delta f$ of the tension due to the longitudinal friction. Estimating
$\delta f\simeq \hat \zeta \avg{\Delta\overline\varrho}/(t L^2)$ from Eq.~(\ref{eq:cg-eom}), we
obtain
\begin{equation}
  \label{eq:delta-f-lelo}
  \delta f\approx  \hat \zeta  \frac{L^2}{\lp}t^{-3/4}\;.
\end{equation}
As in the heuristic discussion of Part~I, we find a diverging tension in the limit
$t\to 0$ for the OPT result, so that the OPT result can only be valid after the
effect of $\delta f$ on the evolution of $\avg{\Delta\overline{\varrho}}(t)$ can be
neglected.  For the particular case of {\lelo}, this time can be determined as
follows. As discussed in Part~I, the stress-free dynamics is at the time $t$
characterized by relaxation of modes with wave length $\lpe(t)\simeq t^{1/4}$. When
$\delta f$ is larger than the critical Euler Buckling force $\lpe(t)^{-2}$
corresponding to the length $\lpe(t)$ we expect that the tension cannot be neglected
for the evolution of $\avg{\Delta\overline{\varrho}}$.  Hence, the above result
$\avg{\Delta\overline{\varrho}}\propto L t^{1/4}/\lp$, obtained from OPT, can only be
valid if
\begin{equation}
  \label{eq:val-cond-lelo}
  \lpe(t)^2\delta f\simeq \frac{L^2}{\lp}t^{-1/4}\ll1 \;,
\end{equation}
i.e., for long enough times
\begin{equation}
  \label{eq:val-cond-lelo-2}
  t\gg t_\star=L^8/\lp^4 \;.
\end{equation}
The time $t_\star(L)$ is obviously not identical with the time
\begin{equation}
  \label{eq:tf-2}
  \totpa=L^2 \lp^{-1} \hat\zeta \fe^{-3/2} \qquad (t\gg\tf)
\end{equation}
it takes for the boundary layer to spread over the filament. To compare them, we
first notice that $\totpa\gg\tf=\fe^{-2}$ implies that the polymer must have been
pre-stretched by a large enough force,
\begin{equation}
  \label{eq:large-force}
  \fe\gg \lp^2/L^4=\epsilon^{-2} L^{-2} \;,
\end{equation}
larger than the pre-stretching force at least necessary to enter
the propagation regime $\tf\ll t\ll t_L^\pa$. By using the estimate
in Eq.~(\ref{eq:large-force}) we can compare $t_\star$ and $\totpa$,
\begin{equation}
  \label{eq:tstar-vs-tf}
  t_\star(L)= \frac{L^2}{\lp}\left(\frac{L^2}{\lp}\right)^3
  \gg\frac{L^2}{\lp} \fe^{-3/2} \simeq \totpa \;.
\end{equation}
It is seen that the time window $\totpa\ll t\ll t_\star$ grows with the
pre-stretching force, which means, in particular, that it describes the limit of an
initially straight polymer.

To determine the physics of this novel regime, we have to solve the equation of
motion for the tension on the finite arc length interval $[0;L]$ using the
approximations developed in Sec.~\ref{sec:asymptotic-growth}.  There, we found that
in the limit $t\gg\tf$ the tension profile of {\lelo} is described by
Eq.~(\ref{eq:pde-letting-loose-long-t-2})
\begin{equation}
  \label{eq:degennes-eq}
  \partial_\sigma^2\varphi=\frac1 4 \varphi^{-3/2}\partial_\tau\varphi\;.
\end{equation}
The right-hand-side represents the time-derivative of the stored
length in the \emph{quasi-static approximation}. Going back to
variables $f$, $s$ and $t$, Eq.~(\ref{eq:degennes-eq}) takes the form
\begin{equation}
  \label{eq:degennes-eq-2}
  \partial_s^2 f=\frac1 4 \frac{\hat \zeta}{\lp}f^{-3/2}\partial_t f \;.
\end{equation}
For the following, we assume that this \emph{quasi-static
approximation} is not only valid in the tension propagation regime for
$\tf\ll t \ll\totpa$, but also for longer times until the OPT
regime begins, $\tf\ll t \ll t_\star$. This is justified a posteriori
in App.~\ref{sec:lelo-app}, where it is shown that the change in
stored length for the solution $f(s,t)$ of
Eq.~(\ref{eq:degennes-eq-2}) can indeed be calculated quasi-statically
for times $t \ll t_\star$. We solve Eq.~(\ref{eq:degennes-eq-2}) for
the boundary conditions
\begin{equation}
  \label{eq:bc}
  f(s=0,t)=0 \;,\;f(s=L,t)=0
\end{equation}
and the initial conditions
\begin{equation}
  \label{eq:ini-cond}
  f(s,0)=\fe  \;,\qquad \text{ for }0<s<L \;.
\end{equation}
In Sec.~\ref{sec:long-time-asymptotics}, we solved the differential equation
Eq.~(\ref{eq:degennes-eq}) with the scaling ansatz $f=\fe \hat \varphi[s/\lpa(t)]$
numerically for the correct initial condition, but ignored the boundary conditions of
at one end by sending $L\to\infty$. In contrast, we now look for a simple solution
that obeys the tension boundary conditions exactly, at the expense of a possible
mismatch with the initial conditions.  To this end, we make the product ansatz
\begin{equation}
  \label{product-ansatz-2}
  f(s,t)=g(t) h(s) \;.
\end{equation}
Separation of variables yields
\begin{equation}
  \label{eq:sep-var}
  \frac{\hat \zeta}{4\lp}g^{-5/2}\partial_t g=C=\sqrt{h}h'' \;.
\end{equation}
Choosing $C=-1/(6L^2)$, for convenience, we find the long time asymptotics
\begin{equation}
  \label{eq:g-eq}
  g(t)\sim\left(\frac{\hat \zeta L^2}{\lp t}\right)^{2/3}\;.
\end{equation}
The spatial part obeys a scaling form $h(s)=\hat h(s/L)$, where $\hat
h$ satisfies the equation
\begin{equation}
  \label{eq:h-eq}
  \partial_\xi^2 \hat h(\xi)=-\frac 1 6 \hat h^{-1/2}  
\end{equation}
with the boundary condition
\begin{equation}
  \label{eq:bc-hat-h}
  \hat h(0)=\hat h(1)=0 \;.
\end{equation}
The analytical solution can be found in a standard way. The main characteristics are
the slope at $\xi=0$,
\begin{equation}
  \label{eq:h-ini-slope}
  \partial_\xi h(\xi)\vert_{\xi=0}=12^{-1/3}\approx 0.4368 \;. 
\end{equation}
and the maximum value of $h$,
\begin{equation}
  \label{eq:hmax}
  h(1/2)=\frac 1 {16} \left( \frac 3 2  \right)^{2/3}\approx
  0.0819 \;.
\end{equation}

\begin{figure}
  \includegraphics[width=\columnwidth]{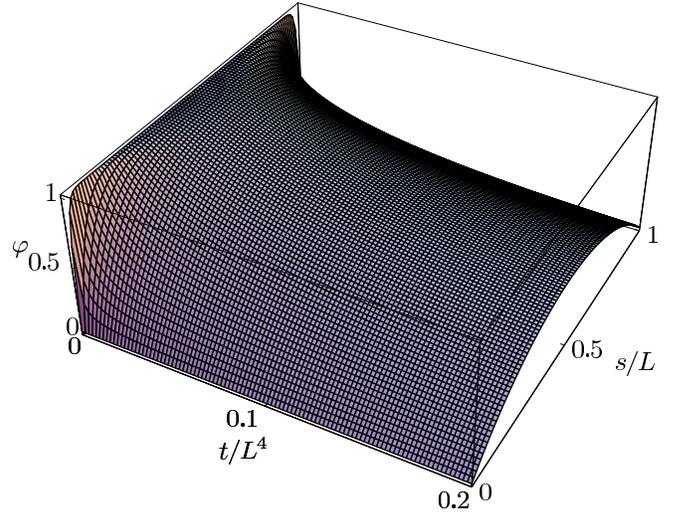}
  \label{fig:lelo-3d}
  \caption{The initial ($t\ll t_\star$) time evolution of {\lelo}. A
    regime of (slow) tension relaxation begins after the sudden change
    in boundary condition has propagated through the filament.}
\end{figure}

Up to now we have investigated the quasi-static approximation in the tension
propagation regime ($t\ll \tfin$) and in the regime of tension relaxation ($t\gg
\tfin$) separately. In order to illustrate the crossover, we have also solved the
corresponding PIDE, Eq.~(\ref{eq:degennes-eq-2}), numerically. The result shown in
Fig.~\ref{fig:lelo-3d} unveils the transient nature of the tension propagation
regime.

\section{Nonlinear response of the projected length}
\label{sec:MSDII}
After having discussed the rich tension dynamics of stiff polymers in detail, we wish
to derive its consequences for pertinent observables in order to make contact with
experiments.  Although the tension may in some situations be monitored directly, see
Sec.~\ref{sec:possible-exps}, a more conventional observable is the longitudinal
extension $R_\pa(t)$ of the polymer, which is defined to be the end-to-end distance
projected onto the longitudinal axis. Tension dynamics strongly affects the nonlinear
response of the projected length, which shall be detailed in the following for the
force protocols {\pull}, {\epull} and {\lelo}.

The average temporal change in the projected length $R_\pa(t)$ is directly related to
the stored length release,
\begin{equation}
  \label{eq:end-end-length}
  \avg{\Delta R_\pa} (t)= -\mint{ds}{0}{L}\avg{\Delta\varrho}(s,t)+o(\epsilon)\;,
\end{equation}
which was already noted in Part~I. As long as modes with wave length on the order of
the total length are irrelevant (i.e.~both ends of the polymer are not correlated),
\begin{equation}
  \label{eq:ends-uncorrelated}
  \lpe(t)\ll L\;,
\end{equation}
half-space solutions for $\avg{\Delta\varrho}$ may be used to evaluate
Eq.~(\ref{eq:end-end-length}). 

Recall from Part~I, that $\avg{\Delta\varrho}$ can be decomposed into a bulk
contribution $\avg{\Delta \overline{\varrho}}$ and a term influenced by the boundary
conditions, which vanishes under a spatial average.  We show in
Sec.~\ref{sec:microscale} that in the cases of hinged and clamped ends, as opposed to
free ends, the contribution of the boundary term to the integral in
Eq.~(\ref{eq:end-end-length}) is subdominant in the limit $t\to 0$.
Nevertheless, these boundary effects represent important corrections that should be
taken into account in any experimental situation (i.e.~with finite $t$) of {\pull}
and {\epull}.

At first, however, let us consider the \emph{bulk} contribution to the end-to-end
distance,
\begin{equation}
  \label{eq:end-end-bulk}
  \avg{\Delta \overline{R}_\pa}(t)\equiv\mint{ds}{0}{L}
  \avg{\Delta\overline{\varrho}}\left[F(s,\tilde t
    \leq t),t\right] \;,
\end{equation}
which is universal in the sense that it does not depend on the boundary conditions of
$\vec r_\pe$. In Eq.~(\ref{eq:end-end-bulk}), and in the following, we neglect contributions
of higher order than $\Delta\varrho=\Ord{\epsilon}$ to the projected end-to-end
distance.   According to Eqs.~(\ref{eq:end-end-bulk},~\ref{eq:cg-eom}),
the bulk change of the end-to-end distance is given by the arclength integral of
the curvature of the time integrated tension $F(s,t)$,
\begin{subequations}
  \begin{align}
    \label{eq:EED-from-tension}
    \avg{\Delta \overline{R}_\pa}(t)&=\hat \zeta^{-1} \mint{ds}{0}{L}F''(s,t)\\
    \label{eq:EED-from-tension-slopes}
    &=\hat \zeta^{-1} \left[  F'(L,t)-F'(0,t) \right]\;.
  \end{align}
\end{subequations}

The predictions are presented separately for the regime of tension propagation ($t\ll
\totpa$) and the regime of OPT ($t\gg t_\star$), where the tension profiles are flat
and time-independent up to sub-leading terms. For weak forces, an explicit expression
for times $t<t_\star$ is given in Sec.~\ref{sec:small-forces-crossover}, which
captures the crossover from tension propagation to the tension saturated OPT regime.
As discussed in Sec.~\ref{sec:homogeneous-relaxation}, {\lelo} turns out to also have
an additional time-window $\totpa< t < t_\star$, whose consequences for the
end-to-end distance is described in
Sec.~\ref{sec:homogeneous-relaxation-e-e-distance}.  Since {\epull} is the only
problem, in which both ends do not behave in the same way, we discuss this
``asymmetrical'' problem separately in Sec.~\ref{sec:e-e-epull}.

\subsection{Tension propagation regime ($t\ll \totpa$)}
\label{sec:unrelaxed-tension}
In the time domain of tension propagation, where the boundary layers are growing in
from both ends ($t\ll \totpa$), we may use the tension profiles for the
semi-infinite (pseudo-) polymer from Sec.~\ref{sec:scaling-forms} (labeled by
$F_\infty$ here) to approximate Eq.~(\ref{eq:EED-from-tension-slopes}) by
\begin{equation}
  \label{eq:EED-from-tension-2}
  \avg{\Delta \overline{R}_\pa}(t)\stackrel{t\ll \totpa}{\sim}-2\hat
  \zeta^{-1} F_\infty'(0,t)
\end{equation}
for scenarios with \emph{two} equally treated ends.  Now, if $t$ falls into a regime
where the tension exhibits scaling,
\begin{equation}
  \label{eq:F-scaling}
  F_\infty(s,t)\propto t^{\alpha+1} \hat F_\infty\left(\frac s {t^z} \right) \;,
\end{equation}
we immediately obtain from Eq.~(\ref{eq:EED-from-tension-2}) the
power-law 
\begin{equation}
  \label{eq:ee-growth-law-principle}
  \avg{\Delta \overline{R}_\pa}(t)\propto - t^{\alpha+1-z} 2 \partial_{\xi} \hat
  F_\infty(\xi=0)
\end{equation}
for the growth of the end-to-end distance. The pre-factors can be calculated for all
cases, because the scaling functions are known. In this way we obtain the list
Tab.~\ref{tab:ww-growth-laws} of growth laws.

\begin{table}
  \caption{Universal bulk contribution $\avg{\Delta\overline{R}_\pa}(t)$ to the dynamic change
    of the end-to-end distance  in the limit
    $t\ll \totpa\ll t_L^\pe$.}
  \label{tab:ww-growth-laws}
  \begin{center}
    \begin{tabular}{c|c|c}
      Problem&$t\ll \tf$&$t\gg \tf$\\ \hline
      {\pull}&${\displaystyle \frac{2^{5/8}}{\Gamma(15/8)}
        \frac{\fe}{\sqrt{ \hat \zeta \lp }}t^{7/8}}$&${\displaystyle
        \frac{4}{\sqrt{3}}
        \left(\frac{2}{\pi}\right)^{1/4}\frac{\fe^{3/4}}{\sqrt{ \hat
            \zeta \lp }}t^{3/4}}$\\
      {\lelo}&$-{\displaystyle \frac{2^{5/8}}{\Gamma(15/8)}
        \frac{\fe}{\sqrt{ \hat \zeta \lp }}t^{7/8}}$&${\displaystyle
        - 2.477\frac{\fe^{1/4}}{\sqrt{ \hat   
            \zeta \lp }}t^{1/2}}$\\
    \end{tabular}
  \end{center}
\end{table}

\subsection{OPT regime ($t\gg t_{\star}$)}
\label{sec:tension-relaxed}
When calculating the release of stored length in OPT the tension profile
$\fopt=f(s=0)=\text{const.}$ is assumed to be stationary and flat, so that
\begin{eqnarray}
  \label{eq:OPT-Rpa}
  &\avg{\Delta \overline{R}_\pa}(t)=-L \avg{\Delta \overline{\varrho}}
  \left[ \fopt,t  \right]\;,\nonumber &\\
   &\qquad\qquad\qquad \text{for }t\gg t_\star \;.&
\end{eqnarray}
In the case of {\lelo}, the quantity $\avg{\Delta \overline{\varrho}}(\fopt=0,t)$ has
been explicitely calculated in Eq.~(\ref{eq:stole-lelo-OPT}).  For {\pull},
$\avg{\Delta \overline{\varrho}}(\fopt=\fe,t)$ can be evaluated from
Eq.~(\ref{eq:change-stored-length}) in a similar straightforward manner, because the
tension is spatially constant. The corresponding growth laws are summarized in
Tab.~\ref{tab:ww-growth-laws-2}.

Compared to the tension propagation regime, the growth laws in the OPT regime are
slowed down, see Tabs.~\ref{tab:ww-growth-laws},~\ref{tab:ww-growth-laws-2}.
Actually, for all cases except {\lelo} the corresponding growth laws obey 
\[ \avg{\Delta \overline{R}_\pa}^{\text{OPT}}\propto \avg{ \Delta
  \overline{R}_\pa}^{\text{MSPT}} t^{-z}\;.\] This can be understood in terms of the
scaling arguments used Part~I.  There, we took tension propagation heuristically into
account by assuming that the stored length release, as given by OPT, is restricted to
the boundary layer of size $\lpa(t)$.  Then, one has 
\begin{equation}
  \label{eq:dr-estimate1}
  \avg{\Delta \overline{R}_\pa}^{\text{OPT}}\simeq -\lpa(t) \avg{\Delta
    \overline{\varrho}}(\fopt,t) \qquad \text{for }t\ll\totpa
\end{equation}
as compared to 
\begin{equation}
  \label{eq:dr-estimate1}
  \avg{\Delta \overline{R}_\pa}^{\text{MSPT}}\simeq - L
\avg{\Delta\overline{\varrho}}(\fopt,t) \qquad \text{for }t\gg t_\star \;.
\end{equation}
This conforms with the heuristic rule noticed above,
\begin{equation}
  \label{eq:OPT-MSPT}
  \avg{\Delta \overline{R}_\pa}^{\text{OPT}}\approx
  \frac{L}{\lpa(t)}
\avg{\Delta \overline{R}_\pa}^{\text{MSPT}}
  \sim t^{\alpha+1-2z}\;.
\end{equation}

\begin{table}
  \caption{Universal bulk contribution $\avg{\Delta\overline{R}_\pa}(t)$ to the dynamic change
    of the end-to-end distance in the limit
    $t_\star\ll t \ll t_L^\pe$.}
  \label{tab:ww-growth-laws-2}
  \begin{center}
    \begin{tabular}{c|c|c}
      Problem&$t\ll \tf$&$t\gg \tf$\\ \hline
      {\pull}&${\displaystyle
        \frac{L
          \fe}{\lp\Gamma(7/4)}\left(\frac{t}{2}\right)^{3/4}}$~\cite{granek:97,gittes-mackintosh:98_pub}
      &${\displaystyle
        \frac{L}{\lp}\sqrt{\frac{2\fe t}{\pi}}}$\\
      {\lelo}&$-{\displaystyle
        \frac{L \fe}{\lp\Gamma(7/4)}\left(\frac{t}{2}\right)^{3/4}}
      $ &${\displaystyle \frac{-2^{3/4}}
        {\Gamma(1/4)} \frac{L}{\lp}t^{1/4}}$\\
    \end{tabular}
  \end{center}
\end{table}

\subsection{{\lelo} in the limit $\totpa\ll t\ll t_{\star}$}
\label{sec:homogeneous-relaxation-e-e-distance}
The intuitive rule in Eq.~(\ref{eq:OPT-MSPT}) fails for {\lelo} in the limit $t\gg\tf$,
which indicates that this is an exceptional scenario.  As discussed in
Sec.~\ref{sec:homogeneous-relaxation}, $t_\star=L^8/\lp^4$ cannot be identified with
$\totpa=L^2 \lp^{-1} \hat\zeta \fe^{-3/2}$ in this case. There exists a time window
$\totpa\ll t \ll t_\star$ that expands in the limit of large forces, $\fe\gg
\lp^2/L^4$.  We have shown that the tension exhibits homogeneous relaxation in this
novel regime.  With the slope of the tension at the ends
\begin{equation}
  \label{eq:lelo-tension-slopes}
  \partial_s f\vert_{s=\left\{  {0 \atop L}\right\}}=\pm \frac 1 {16} \left( \frac 3 2
  \right)^{2/3} \frac{1}{L}\left(\frac{\hat \zeta L^2}{\lp t}\right)^{2/3}
\end{equation}
the growth law 
\begin{equation}
  \label{eq:homogenous-lelo-3}
  \avg{\Delta \overline{R}_\pa}(t)\simeq - 18^{1/3} \left(\frac{L t}{\hat \zeta
      \lp^2}\right)^{1/3} 
  \qquad (\lelo)  
\end{equation}
follows from Eq.~(\ref{eq:EED-from-tension-slopes}). We expect the growth law
$\avg{\Delta \overline{R}_\pa}\propto t^{1/3}$ during homogeneous tension relaxation to hold even
for chains with $L\gg\ell_p$. The example of retracting DNA will be discussed as an
experimental outlook in Sec.~\ref{sec:possible-exps}. The exponent $1/3$ coincides
with that obtained by an adiabatic application of the stationary force-extension
relation~\cite{bustamante-marko-siggia-smith:94} to a
``frictionless''~\cite{bhobot-wiggins-granek:04} polymer with attached beads at its
ends~\cite{Hallatschek:F:K::94:p077804:2005}.

For times $t\gg t_\star=L^8/\lp^4$ the growth law in Eq.~(\ref{eq:homogenous-lelo-3})
crosses over to the one noted in Tab.~\ref{tab:ww-growth-laws-2}. Interestingly, both
growth laws appearing for $t\gg \totpa$ are independent of the initial tension $\fe$.
In both cases, the initial conditions are completely ``forgotten'' once the tension
has propagated through the whole polymer. An overview over the time scales separating
the diverse regimes for {\lelo} (as compared to {\pull}) will be given in
Sec.~\ref{sec:lelo-vs-pull}.

\subsection{{\pull} and {\lelo} for small forces}
\label{sec:small-forces-crossover}
\begin{figure}
  \includegraphics[width=\columnwidth]{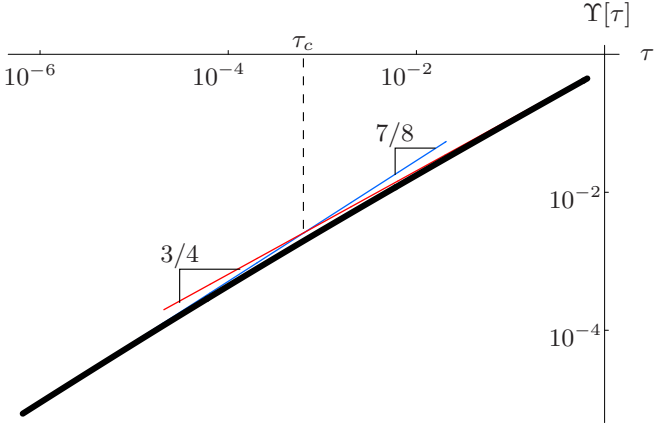}
  \caption{The scaling function $\Upsilon(\tau)$ in a double logarithmic plot. The
    crossover of the scaling behavior is close to the time $\tau_c =6.55\times
    10^{-4}$ where the asymptotic short-(blue) and longtime (red) asymptotics
    cross.}
  \label{fig:pulling-finite-polymer-weak-forces}
\end{figure}

Provided the external force is smaller than the critical Euler buckling force of the
polymer, $\fe< L^{-2}$, the crossover time $\tf$ exceeds the terminal relaxation time
$\totpe$, hence the linearized PIDE, Eq.~(\ref{eq:pide-linearized-lt-1}), applies
throughout the contour relaxation. The linearity allows to solve the PIDE for a
polymer of finite length, and we obtain an analytic description of the crossover
between the asymptotic power laws $\Delta\overline{R}_\pa(t)\propto t^{7/8}$ for
$t\ll t_L^\pa$ and $\Delta\overline{R}_\pa(t)\propto t^{3/4}$ for $t_\star\ll t \ll
t_L^\pe$ (cf.~Tabs.~\ref{tab:ww-growth-laws},~\ref{tab:ww-growth-laws-2}). To this
end, let us first reinstall original units (which are better adapted for the
present purpose) into the linearized PIDE, Eq.~(\ref{eq:pide-linearized-lt-1}),
\begin{equation}
  \label{eq:PIDE-linear-units}
  \partial_s^2 F(s,z)=\frac{z^{1/4} \hat \zeta}{2^{3/4} \lp} F(s,z)\;,
\end{equation}
where $F(s,z)$ is the Laplace transformation of the integrated tension,
\begin{equation}
  \label{eq:Laplace trafo}
  F(s,z)\equiv \mint{dt}{0}{\infty}e^{-z t}F(s,t)\;.
\end{equation}
For the boundary conditions of {\pull}, Eqs.~(\ref{eq:bc-infty},
\ref{eq:bc-origin-1}), the solution to Eq.~(\ref{eq:PIDE-linear-units}) is given by
\begin{equation}
  \label{eq:weak-pulling-finite-polymer}
  F(s,z)=\fe z^{-2}\frac{\cosh\left[(\hat \zeta/\lp)^{1/2} (z/8)^{1/8} (s-L/2)
    \right]} {\cosh\left[(\hat \zeta/\lp)^{1/2} (z/8)^{1/8} L/2  \right]}
\end{equation}
This implies a growth of the end-to-end distance of
\begin{equation}
  \label{eq:eed-weak-forces-finitep-laplace}
  \begin{split}
    \avg{\Delta \overline R_\pa}(z)&=-2\hat \zeta^{-1} \partial_s F|_{s=0} \\
    &=\frac{2^{5/8}\fe}{z^{15/8} (\hat \zeta\lp)^{1/2}}\tanh\left[ \left( \frac{\hat
          \zeta}{\lp} \right)^{1/2}\left( \frac{z}{8} \right)^{1/8} \frac{L}{2}
    \right]\;.
  \end{split}
\end{equation}
The Laplace back-transform of Eq.~(\ref{eq:eed-weak-forces-finite-polymer}) takes
the form 
\begin{equation}
  \label{eq:eed-weak-forces-finite-polymer}
  \begin{split}
  \avg{\Delta \overline R_\pa}(t)&=\mint{\frac{dz}{2\pi
      i}}{-i\infty+\epsilon}{i\infty+\epsilon} e^{z t}   \avg{\Delta \overline
    R_\pa}(t)\\
  &=\frac{\fe \hat \zeta^3 L^7}{\lp^4} \Upsilon\left[ \left( \frac{t}{L^4}  \right)
    \left( \frac{\lp}{\hat \zeta L}  \right)^4 \right]
  \end{split}
\end{equation}
where $\Upsilon(\tau)$ is a scaling function, given by
\begin{equation}
  \label{eq:Upsilon}
  \begin{split}
    \Upsilon(\tau)&= \frac{2^{5/8}}{\pi}\mint{ds}{0}{\infty}\frac{e^{-x
        \tau}-1}{x^{15/8}}\\
    &\times \Im\left[ e^{-i\frac 78 \pi} \tanh\left( 2^{-11/8}x^{1/8}e^{i
          \pi/8} \right) \right]
  \end{split}
\end{equation}
We have depicted $\Upsilon(\tau)$ in a double logarithmic plot in
Fig.~\ref{fig:pulling-finite-polymer-weak-forces} together with the corresponding
asymptotics from Tabs.~\ref{tab:ww-growth-laws},~\ref{tab:ww-growth-laws-2}. Note
that the time $\tau_c\approx 6.55 \times 10^{-4}$ where the asymptotic lines cross is
a good indication for the crossover occurring in the exact solution. Assuming that
this also holds quite generally, it is possible to obtain estimations of when the
crossover between linear and nonlinear regimes should occur in the natural time units
$\tf$. To this end, one simple equates the corresponding asymptotic power laws
(including the exact pre-factors) for the growth of the end-to-end distance.  As for
the present case with $\tau_c\approx 6.55 \times 10^{-4}$, these crossover times are
typically not of order one in natural units because of the numerical proximity of the
exponents of the asymptotic power laws, which are $7/8$ and $3/4$ in the present
case. In a given experimental situation, one should therefore check carefully which
regime is expected by comparing experimental time scales with these unusual crossover
times.

\subsection{FDT and linear response}
\label{sec:tens-fluct}
According to the fluctuation-dissipation theorem
(FDT~\cite{kubo83bookI,kubo91bookII}), the fluctuations in the end-to-end distance
should be related to the linear response of the polymer. It is tempting to interpret
the linear short-time regime, where $\avg{\Delta \overline{R}}\simeq \fe t^{7/8}(\hat
\zeta \lp)^{-1/2}$, as linear response in the usual sense, in which case the
fluctuations of the end-to-end distance should, according to the FDT, scale as
$\avg{\Delta \overline{R}^2}\simeq t^{7/8}{\hat \zeta}^{-1/2} \lp^{-3/2}$ in
equilibrium.  However, fluctuations of this strength are inconsistent with a
deterministic tension dynamics, because they generate friction forces over a scale
$\lpa(t)\simeq t^{7/8}(\lp/\hat \zeta)^{1/2}$ and thus imply tension fluctuations of
magnitude
\begin{equation}
  \label{eq:tens-fluct}
 \delta f\simeq \hat \zeta  \lpa(t) \avg{\Delta  \overline{R}}/t\simeq
 (\zeta/\lp)^{1/4} t^{-7/16}
\end{equation}
Tension fluctuations exceed the applied force in magnitude at any given time for
small enough external forces. Hence, in the limit $\fe\to 0$ while $\epsilon\ll1$ is
fixed, the tension cannot be considered as a deterministic quantity. Recall however,
that our MSPT analysis in Part~I was based on the limit $\epsilon\to 0$ while $\fe$
is fixed, and only in this limit the self-averaging argument of Part~I applies.
Extending our results to the usual linear response limit corresponds to the
uncontrolled approximation $f \rpe'^2\to f\avg{\rpe'^2}$. To analyze this limit more
carefully, one has to solve the stochastic PIDE that obtained in Part~I before the
self-averaging argument was employed.  However, since the $7/8-$scaling of the
fluctuations has already been confirmed in simulations~\cite{everaers-Maggs:99}, we
expect that such a more rigorous analysis would yield the same scaling but a
pre-factor different from the one of the deterministic short-time law
(Tab.~\ref{tab:ww-growth-laws}).

\subsection{{\lelo} versus {\pull} (overview)}
\label{sec:lelo-vs-pull}
\begin{figure}
  \includegraphics[width=\columnwidth]
  {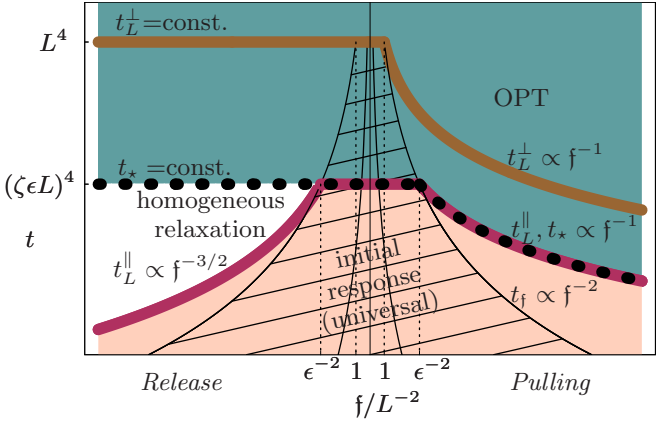}
  \caption{Characteristic times (logarithmic scale) for {\pull} and {\lelo} against
    the applied external force $\fe$ (linear scale). The time $t_\star$ (stars)
    separates regions where ordinary perturbation theory (OPT) applies (dark shaded)
    from regions (light shaded) of linear (hatched) and nonlinear tension propagation
    and from homogeneous tension relaxation (white). Whereas longitudinal friction is
    negligible for $t>t_\star$, it limits the dynamics for $t<t_\star$. The innermost
    funnel-shaped region indicates the regime where the tension fluctuations are
    important (defined by $f<\delta f$ with $\delta f$ given by
    Eq.~(\ref{eq:tens-fluct})).}
   \label{fig:lelo-vs-pull}
\end{figure}

The diverse regimes and their range of validity are summarized in
Fig.~\ref{fig:lelo-vs-pull} for the {\pull} and the complementary
{\lelo} problem. It depicts the crossover time scales as a function of
the externally applied tension $\fe$. The line $\tf\simeq \fe^{-2}$
separates ``linear'' from ``nonlinear'' behavior. The symmetry of the
graph for $t<\tf$ with respect to the $\fe=0$-axis indicates that the
scenario of {\pull} can indeed be considered as the inverse scenario
of {\lelo} for weak forces $f\ll f_c$, or, more generally, on short
times. This symmetry is lost in the nonlinear regime. The growing
importance of uniform tension relaxation for {\lelo} with increasing
initial tension $\fe$ becomes particularly apparent on the log-scale
of the figure. Due to the particular choice of units ($t/L^4$ and
$f/L^{-2}$) the regimes of nonlinear growth of the boundary layers
appear relatively narrow in Fig.~\ref{fig:lelo-vs-pull}. Which of the
various regimes will pre-dominantly be observed in measurements
actually depends strongly on the ratio $L/\lp$ and on the
experimentally accessible time scales.

\subsection{\epull}
\label{sec:e-e-epull} {\epull} was excluded from the preceeding discussion because it
is the only ``asymmetric'' problem, as both ends of the polymer behave differently.
In the tension propagation regime ($t\ll\totpa$), the left end is constrained to move
with constant velocity, while the right end experiences no driving force.  Hence, we
have
\begin{equation}
  \label{eq:epull-growth-law}
  \avg{\Delta \overline{R}_\pa}(t)=v t \;,\qquad \text{for } t\ll\totpa\;.
\end{equation}
After the boundary layer of non-zero tension has reached the free end, $t\gg \totpa$,
the right end starts to move. Then, the tension profile becomes linear as for a
straight rod dragged through a viscous solvent. The further contour relaxation is up
to pre-factors identical to the {\pull} problem for $t\gg \totpa\approx t_\star$.

\subsection{\push}
\label{sec:pushing}

For completeness, we mention the scenario of a filament being compressed by external
longitudinal forces. This scenario has some subtleties.  {\push} increases the stored
length exponentially for $t\gg \tf$ by virtue of the Euler Buckling instability and
generates a situation where the weakly bending approximation is not valid anymore.
Then, hairpins are generated~\cite{ranjith:02} and for $t\gg t_f$ the rigidly
oriented driving forces \emph{pull} on those hair pins. Our theory is only applicable
at short times. For $t\ll t_f$ the response of the system is linear in the driving
force, irrespective of the sign.

\subsection{Boundary effects}
\label{sec:microscale}
Up to now, we only discussed the change $\avg{\Delta \overline{R}_\pa}(t)$ in
projected end-to-end distance corresponding to the change
$\avg{\Delta\overline{\varrho}}$ of the stored length in the bulk. For a hinged,
respectively, clamped semi-infinite polymer, the as yet missing boundary
contribution $X^{h/c}(t)\equiv\avg{\Delta R^{h/c}_\pa}-\avg{\Delta\overline{R}_\pa}$
is given by
\begin{eqnarray}
  \label{eq:bc-contri-1}
   && X^{h/c}(t)=\pm2\mint{ds}{0}{\infty}\mint{\frac{dq}{\pi\lp}}{0}{\infty}
  \left\{\frac{e^{-2q^2[q^2 t+ F(s,t)]}-1}{q^2+f_<}\right. \nonumber\\ 
    &&+\left. 2 
    q^2\mint{d\tilde t}{0}{t}e^{-2q^2\left[q^2(t-\tilde
        t)+F(s,t)-F(s,\tilde t)\right]}\right\}\cos(2 q s) \;.\nonumber\\
\end{eqnarray}
As discussed in Part~I, the boundary dependent term
of the stored length decays on a length scale of $\Ord{1}$ due to the cosine factor.
Since the tension decays on a much larger length scale of order
$\Ord{\epsilon^{-1/2}}$, it is permissible to use the (integrated) tension at $s=0$
to evaluate the arclength integral in Eq.~(\ref{eq:bc-contri-1}).
 
Upon replacing $F(s,t)\to F(0,t)$ and using 
\[\mint{ds}{0}{\infty}\cos(2 q s)=(\pi/2)\delta(q)\;,\] the integral in Eq.~(\ref{eq:bc-contri-1}) can be evaluated,
\begin{equation}
  \label{eq:Rpa-boundary}
  X^{h/c}(t)=\pm\lim_{q\to 0}\frac 2 {\lp} \frac{q^2(q^2 t +F(0,t))}{q^2+f_<}\;,
\end{equation}
which vanishes, unless
\begin{equation}
  \label{eq:Rpa-boundary2}
  f_<=0 \quad \Rightarrow \quad X^{h/c}(t)=\pm\frac {2F(0,t)} {\lp}\;.
\end{equation}
On the semi-infinite arclength interval, these boundary contributions are therefore
nonzero only for {\pull} and {\epull}, and are summarized in
Tab.~\ref{tab:tab:bc-effects}. Note that, the boundary effects of clamped (hinged)
ends tend to reduce (increase) the longitudinal response of the polymer in comparison
to the bulk response.  This may be explained as follows. Close to a clamped end, a
polymer is more stretched out than in the bulk because $\vec r_\pe'^2$ is constraint
to approach zero at the end. As a consequence, the end portion of the polymer is less
able to store or release excess length. For hinged ends, the boundary conditions act
just in the reverse direction.

\begin{table}
  \caption{Boundary contribution to the change $\avg{\Delta R^{h/c}_\pa}(t)$ of the end-to-end distance for
    hinged, respectively, clamped ends in the limit $t_L^\pe \ll t$.} 
  \label{tab:tab:bc-effects}
  \begin{tabular}{c|c}
    Problem&$\avg{\Delta R^{h/c}_\pa}(t)-\avg{\Delta \overline{R}_\pa}(t)$\\ \hline
    {\pull}&$\pm 2t \fe/\lp$\\
    {\epull}&$\pm 2t \fe/\lp$\\
    {\lelo}&$\left\{ {\pm 2t \fe/\lp\text{, for }f_<\ll L^{-2}  \atop 
        0 \text{, for }f_<\gg L^{-2}}
    \right.$ \\
  \end{tabular}
\end{table}

The strict vanishing of the boundary term for any finite $f_<$ and the discontinuity
at $f_<=0$ is a consequence of the assumed infinite half-space. For a polymer of
finite length, it turns out that the boundary term approaches zero for
pre-stretching forces larger than the critical Euler-Buckling force, $f_<\gg
f_c\equiv L^{-2}$. This can be seen by studying the finite integral
$\mint{ds}{0}{L}(\avg{\Delta \varrho^{h/c}}(s,t)-\avg{\Delta \overline{\varrho}}(s,t))$. For
forces $f_<\ll f_c$, the integrand saturates at a plateau of magnitude
$\Ord{f_<^{3/2}t/\lp }$ for $\lpe(t)<s<f_<^{-1/2}$ before it finally decays to zero.
Within the semi-infinite integral, the integral over this long plateau cancels the
contribution stemming from $s<\lpe(t)$, as required by Eq.~(\ref{eq:Rpa-boundary}).
However, for $L\ll f_<^{-1/2}$ the contribution from the plateau may be neglected. As
a consequence, the value of the integral is for $f_<\ll f_c$ given by $\pm 2f_<
t/\lp$.  This asymptotic behavior is important for {\lelo}, as noted in
Tab.~\ref{tab:tab:bc-effects}.

Upon comparing Tab.~\ref{tab:tab:bc-effects} with Tabs.~\ref{tab:ww-growth-laws},
\ref{tab:ww-growth-laws-2}, one may think that boundary effects are always
subdominant in the short time limit.  However, our calculations were specialized to
hinged or clamped boundary conditions.  In many experimental situations, one has to
deal with \emph{free} boundary conditions.  Somewhat tedious but rigorously, these
boundary conditions can be taken into account by means of the correct susceptibility,
which can only be given in terms of an integral. Here, we discuss effects related to
free boundary conditions on a heuristic basis and show that they generate a
\emph{dominant} contribution to the change in the end-to-end distance for {\pull},
which is proportional to $t^{3/4}$.

The argument is based on the observation, that external forces that act in the
longitudinal direction ($\pa$) while the polymer is free, automatically introduce
small (of order $\Ord{\epsilon^{1/2}}$) transverse forces at the ends.  This follows
from the force balance equation of a semiflexible polymer (cf.~Part~I)
\begin{equation}
  \label{eq:force-balance}
   \kappa\, \vec{r}''' +\fel =  f\, \vec{r}' \;,
\end{equation}
where $\fel(s)$ is the elastic force acting at arc length $s$. At the ends, where
$\fel$points in longitudinal direction to cancel the external pulling force, one obtains
from the projection of Eq.~(\ref{eq:force-balance}) onto the transverse axis
\begin{equation}
  \label{eq:perp-endforces}
  \vec r_\pe'''=f \vec r_\pe'=\fe \vec r_\pe'+\Ord{\epsilon} \qquad \text{(at the ends)} \;.
\end{equation}
Thus, $\vec r_\pe'''$ is typically non-zero at the ends because the slope $\vec
r'=\Ord{\epsilon^{1/2}}$ fluctuates.  This, however, corresponds to a transverse
force at the end, which results in a transverse deformation. The corresponding bulge
of contour is only visible on the micro-scale because it spreads with the transverse
correlation length $\lpe(t)\simeq t^{1/4}$.  Nevertheless this deformation may
dominate the growth of the end-to-end distance, as is demonstrated for {\pull}: With
the transverse bulk susceptibility scaling as $\overline{\chi}(t)\simeq t^{-1/4}$ we
estimate the magnitude of transverse deformation $\Delta r_\pe(s=0,t)\equiv r_\pe
(0,t)-r_\pe(0,0)$ induced by a transverse force of magnitude
$\ord{\fe\epsilon^{1/2}}$ by 
\begin{equation}
  \label{eq:tdispl-estimate}
  \ord{\avg{\abs{\Delta r_\pe(0,t)}}}\simeq \fe
  \epsilon^{1/2} t \, \overline{\chi}(t)\simeq \epsilon^{1/2} t^{3/4}\fe \;.
\end{equation}
The displacement of the end couples to the projected length because of the mismatch
of the end-tangent with the $\pa$-axis by a small angle of typical magnitude
$\avg{\abs{r_\pe'(s=0)}}\simeq \ord{\epsilon^{1/2}}$. The expected growth of
end-to-end distance due to this effect is therefore estimated by
\begin{equation}
  \label{eq:micro-fluctuations}
  \begin{split}
    &\avg{\Delta R^{h/c}_\pa}(t)-\avg{\Delta
      \overline{R}_\pa}(t)\simeq\\&
    \avg{\abs{r_\pe'(s=0)}} \times\avg{\abs{\Delta
      r_\pe(0,t)}}\simeq\epsilon t^{3/4}
    \fe=\frac{L}{\lp}t^{3/4}\fe  \;.
  \end{split}
\end{equation}
This dominates over the growth law of the bulk, which scales like
$t^{7/8}/\sqrt{\lp}$ on short times. The only way to avoid the outlined effect is to
apply the external force strictly tangentially to the end-tangents, which is however
somewhat unrealistic.  The same problem will experimentally arise in the short time
limit of {\lelo} if the pre-stretching force was not applied strictly tangentially.
However, the long-time limits are unaffected by this subtlety because the bulk
contributions dominate over contributions from the end.

To our knowledge, these end-effects have so far masked the subdominant
$t^{7/8}$-contribution in experiments that monitored the time-dependent end-to-end
distance (we note that in Ref.~\cite{legoff-amblard-furst:02} the $7/8$-scaling is
inferred from a corresponding scaling of the measured shear modulus of an active
gel).  As we outline in Sec.~\ref{sec:smfs}, force spectroscopy, on the contrary, may
allow to measure the tension dynamics, which is itself truly \emph{independent} of
the boundary conditions imposed on the transverse displacements.

\section{Suggestions for experiments}
\label{sec:possible-exps}
While many experiments have been done concerning flexible polymers in external force
fields~\cite{perkins-smith-chu:97,quake-babcock-chu:97}, the available measurements
on driven stiff or pre-stretched polymers is not sufficient to verify our
predictions.  Most of these experiments have monitored the transverse and
longitudinal response on intermediate times where OPT is valid
(e.g.~\cite{legoff-hallatschek-frey:02}). Investigations concerning the longitudinal
short-time dynamics are
scarce~\cite{bhobot-wiggins-granek:04,legoff-amblard-furst:02,maier-seifert-raedler:02}.
In the following, we propose several assays that might be able to fill this gap.

To facilitate the application of our predictions, we reintroduce the parameters
$\kappa$ and $\zeta$ for the following.

\subsection{Strongly stretched DNA}
\label{sec:dna-exp}

The experimental verification of most of our results requires stiff
polymers with a total length much smaller than their persistence
length.  A remarkable exception is {\lelo} in the regime of
homogeneous tension relaxation, which was discussed in
Sec.~\ref{sec:homogeneous-relaxation}. For polymers with $L\gg\lp$,
like a typical DNA molecule, this novel regime appears if the
pre-stretching force obeys
\begin{equation}
  \label{eq:lelo-exp-large-forces}
  \fe\gg \fe_{\lp}\equiv \kappa \lp^{-2}=\frac{k_B T}{\lp} \;, 
\end{equation}
where $\fe_{\lp}$ is the Euler buckling force corresponding to the
buckling length $\lp$. For forces much larger than the cross-over
force $\fe_{\lp}$, the polymer may be considered as weakly bending.
For those long, but strongly stretched polymers our analysis of
{\lelo} predicts the following.

After the stretching force has been released the tension first
propagates through the filament. This
regime~\cite{brochard-buguin-de_gennes:99,Hallatschek:F:K::94:p077804:2005}
ends at the time
\begin{equation}
  \label{eq:lelo-exp-crossover}
  \totpa=\frac{\zeta_\pa L^2 \lp}{k_b T}\left( \frac{\fe}{\fe_{\lp}}
  \right)^{-3/2}\equiv \tau_{R} \left( \frac{\fe}{\fe_{\lp}}  \right)^{-3/2}
\end{equation}
when the tension has to propagated through the filament. Here, $\tau_{R}\equiv
\zeta_\pa L^2 \lp /(k_B T)$ scales like the longest relaxation time of a Rouse chain
with segment length $\lp$. The characteristic time $\totpa\ll \tau_{R}$ marks the
cross-over to a regime where the tension profile is roughly parabolic and slowly
decays according to the power law $f(t)\propto t^{-2/3}$. This is associated with the
projected length $R_\pa$ growing like
\begin{equation}
  \label{eq:lelo-exp-growth-law}
  \avg{\Delta R_\pa(t)}\equiv \avg{R_\pa(t)-R_\pa(0)}=
  18^{1/3} L \left( \frac{t}{\tau_R}
  \right)^{1/3} \;. 
\end{equation}

The above analysis strictly holds for the portion of the polymer that
stays weakly bending. This is not the case at the boundaries, for
which we refer to existing theories. According to the stem-flower
model of Refs.~\cite{brochard:93,brochard:95,manneville:96} the
boundaries will develop in time $t$ a ``flower'' of arc length
\begin{equation}
  \label{eq:flower-length}
  \lflower(t)\simeq L \left( \frac{t}{\tau_{R}}  \right)^{1/2}  \;,
\end{equation}
thereby reducing the end-to-end distance by an amount of the same order of
magnitude as $\lflower$ itself. It is seen, that for $t\ll \tau_R$ the shrinkage of
the end-to-end distance due to the flower is much smaller than that due to the
weakly bending part of the polymer (the stem), Eq.~(\ref{eq:lelo-exp-growth-law}).

Thus, the evolution of the end-to-end distance should be described
by Eq.~(\ref{eq:lelo-exp-growth-law}) even for flexible polymers if
the pre-stretching force is large enough.  Since DNA can be stretched
by very large forces without un-zipping or destroying the covalent
bonds, we think that the scaling $\avg{\Delta R(t)}\propto t^{1/3}$
should be  visible in a {\lelo}-experiment with DNA. The relevant
quantities in an experiment with $\lambda$-phage DNA (as in
Ref.~\cite{maier-seifert-raedler:02}) in aqueous solution would be
\begin{subequations}
  \label{eq:lelo-exp-numbers}
  \begin{align}
    \lp&\approx 50\text{nm} \nonumber \\
    L&\approx 20 \mu\text{m}  \nonumber\\
    a&\approx 2 \mu\text{m}  \qquad \text{(thickness)}\nonumber\\
    \zeta_\pe&\approx 4 \pi \eta/\log(L/a)\approx 1.3 \times 10^{-3}
    \text{Pa s}  \nonumber\\
    \fe_{\lp}&\approx 0.08 \text{pN}  \nonumber\\
    \tau_R&\approx 7 \text{s} \;.  \nonumber
  \end{align}
\end{subequations}
Though it might require extreme conditions to reach the asymptotic limit in
Eq.~(\ref{eq:lelo-exp-growth-law}), any experiment with finite pre-stretching forces
larger than $f_c$ should be suitable for deciding the question whether or not the
stem dominates the retraction.  One would then compare the data with a numerical
solution~\cite{obermayer-kroy-frey-hallatschek:tbp} of the governing equation,
Eq.~(\ref{eq:cg-eom}), which describes the stress relaxation in stiff worm-like
chains.

\subsection{Single molecule force spectroscopy}
\label{sec:smfs}

\subsubsection{{\epull}}

\begin{figure}
  \centerline{\includegraphics[width=.7\columnwidth]
    {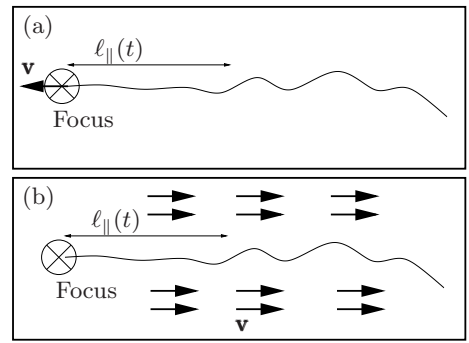}}
  \caption{Two possible realizations of {\epull}. In (a) the polymer's left end is
    being dragged through the solvent with constant velocity $\vec v$, whereas in (b)
    the optical tweezer is immobile while the solvent flows with constant velocity
    $\vec v$. In both experiments, the length $\lpa(t)$ of the boundary layer is
    derived from the pulling force of the tweezer, which can be inferred from the
    displacement of the end within the focus of the tweezer.}\label{fig:tweezer}
\end{figure}

As we already mentioned Sec.~\ref{sec:tensprop-summary}, {\epull} offers the possibility
to measure the tension propagation by a force measurement.  The experimental idea is
illustrated in Fig.~\ref{fig:tweezer} (a). The polymer's left end is suddenly pulled,
e.g., by an optical tweezer whose focus moves with constant velocity. As a
consequence the pulled end follows the laser beam with (almost~\footnote{Due to the
  finite stiffness of the harmonic Laser potential, the trapped end will not be moved
  with \emph{exactly} the same velocity as the trap during a transient, in which the
  end approaches its steady state location within the Laser potential.})  constant
velocity. By measuring the deflection of the trapped end from the center of the beam
one can, in principle, extract the pulling force $\fe(t)$ and thus the size
$\lpa(t)\simeq \fe(t)/(\hat \zeta v)$ of the boundary layer.  However, since the
laser beam is moving, it might represent some problems to dynamically extract the
deflection.

A solution to the latter problem is suggested by the following Gedanken experiment.
Consider the above realization of {\epull} in the coordinate frame co-moving with
the left tip, as illustrated in Fig.~\ref{fig:tweezer}(b). Then, it seems as if the
bulk of the polymer was dragged by a homogeneous force field to the right while the
left end is held fixed by the optical tweezer, just as if the optical tweezer was
spatially fixed while the solvent was homogeneously flowing to the right. In fact,
from the polymer's perspective there is no difference between both experiments. Thus,
we propose to graft one end of a polymer by a tweezer or by the cantilever of an
atomic force microscope.  Then, a homogeneous force field (electric field or fluid
flow) is suddenly turned on that pulls the bulk of the polymer to the right.  The
deflection of the tip gives the pulling force and hence the length of the boundary
layer.

\subsubsection{Onset of the nonlinear regime}

One would like to estimate typical time sales for the diverse regimes
of tension propagation and relaxation. As we have seen above, most of
the time-scales crucially depend on the total length of the polymer,
e.g., $\lpa(t)\propto L^8$ in the linear regime. This high tunability
is of experimental advantage because one can adjust the setup to the
observable time scales. On the other hand, it forbids to give
\emph{typical} time-scales for those quantities.

\begin{table}
  \caption{Threshold values $I_c=\sqrt{\kappa \zeta_\pe}$ for diverse biopolymers.}
  \label{tab:threshold-ic}
  \begin{center}
    \begin{tabular}{c|c}
      Polymer&$I_c$\\ \hline
      microtubuli & $1.6 \times 10^2$ pN\,ms$^{1/2}$ \\ 
      F--actin & $10$ pN\,ms$^{1/2}$\\
      Intermediate fil.s & $4$ pN\,ms$^{1/2}$ \\
      DNA & $0.9$ pN\, ms$^{1/2}$  \\
    \end{tabular}
  \end{center}
\end{table}

However, there are characteristic quantities that do not depend on the
length of the polymer. The probably most interesting one is the
threshold to the nonlinear regime: when the product of driving force
and the square root of the applied time exceeds a certain value $I_c$
the polymer response becomes nonlinear,
\begin{equation}
  \label{eq:nonlinear regime}
  \fe \sqrt{t} \gg I_c\equiv\sqrt{\kappa \zeta_\pe}=\sqrt{k_B T \lp \zeta_\pe}
    \;.
\end{equation}
With persistence lengths of $\lp=7$mm, $17\mu$m, $2\mu$m, $50$nm for
microtubuli\cite{mickey-howard:95,FrancescoPampaloni07052006},
F-actin~\cite{legoff-hallatschek-frey:02}, intermediate filaments and
DNA~\cite{bustamante-marko-siggia-smith:94}, respectively, and corresponding friction
coefficients $\zeta_\pe\approx 4 \pi \eta /\ln(\lp/a)$ (transverse friction
coefficient per length of a rod of length $\lp$), we have evaluated $I_c$ for some
common biopolymers, see Tab.~\ref{tab:threshold-ic}.  Those values can be used to
decide whether the response of a given biopolymer under a ``typical'' time-dependent
external longitudinal force is predominantly nonlinear or linear.  During a power
stroke, for instance, the molecular motor myosin exerts a force of about $5$pN on an
actin filament during a time of roughly $1$ms (cycle time of the power stroke).
Hence, the impulse of about $0.5 I_c$ is somewhat smaller than $I_c$ for actin, so
that the actin response should be linear.

\section{Summary}
\label{cha:Summary}

In this paper, we have studied the tension dynamics of a weakly bending semiflexible
polymer in a viscous fluid theoretically. Starting from the coarse-grained equation
of motion for the tension, Eq.~(\ref{eq:pide}), we elaborated the non-linear
longitudinal dynamic response to various external perturbations (mechanical
excitations, hydrodynamic flows, electrical fields \dots) that can be represented as
sudden changes of boundary conditions.  For the various scenarios we identified
two-parameter scaling forms, that capture the crossover from linear to nonlinear
tension dynamics.  In the limit of large and small arguments, where the equilibrium
structure of the polymer is self-affine, they were shown to reduce to one-parameter
scaling forms, which could be calculated analytically in most cases.  The growth law
$\lpa(t)\sim t^z$ of the tension profiles could be inferred from the scaling
variables of the respective scenarios. This enabled us to develop a unified theory of
tension propagation. Not only does it contain all cases (correctly) studied in the
literature so far.  It also identifies their ranges of validity and provides new
predictions.  The recovered known results and our new predictions are summarized in
Figs.~\ref{fig:lf},~\ref{fig:lelo-vs-pull} and
Tabs.~\ref{tab:lf-growth-laws},~\ref{tab:ww-growth-laws} and
\ref{tab:ww-growth-laws-2}.  

Various dynamic regimes should be well realizable for certain biopolymers.  A novel
regime of homogeneous tension relaxation is a particularly remarkable result from the
experimental point of view (Sec.~\ref{sec:dna-exp}).  In contrast to previous
expectations, this new regime is predicted to dominate the relaxation of strongly
stretched DNA. Moreover, it is an intriguing question, whether the tension
propagation laws $\ell_\|(t)$ govern mechanical signal transduction through the
cytoskeleton~\cite{shankar-pasquali-morse:2002,GardelVCBW03}. We expect that the
force spectroscopical methods, proposed in Sec.~\ref{sec:smfs}, might be helpful to
answer these questions.

Inclusion of hydrodynamic interactions merely produce logarithmic corrections but
would give rise to more interesting effects for polymerized membranes to which our
discussion could be generalized with otherwise little change. Other natural
generalizations including the transverse nonlinear response of
polymers~\cite{obermayer-hallatschek2:tbp}, quenches in the persistence
length~\cite{obermayer-kroy-frey-hallatschek3:tbp} and more complex force
protocols~\cite{obermayer-hallatschek-frey-kroy:tbp} are currently also under
investigation.

\section{Acknowledgments}
\label{sec:ack}
It is a pleasure to acknowledge helpful conversations with Benedikt Obermayer, who,
in addition, corrected several pre-factors.  This research was supported by the
German Academic Exchange Service (DAAD) through a fellowship within the
Postdoc-Program (OH) and by the Deutsche Forschungsgemeinschaft through grant no.~Ha
5163/1 (OH) and SFB 486 (EF).

\appendix

\begin{table}[b]
  \caption{Some important notations}
  \label{tab:common-notation}
  \begin{ruledtabular}
    \begin{tabular}{c|l}
      Symbol(s) & General Meaning \\ \hline
      $L$ & total length of the worm-like chain \\
      $\kappa$ & bending stiffness \\
      $\epsilon$  & small parameter, defined such that $\vec r_\pe'^2=\Ord{\epsilon}$    \\
      $\vec r_\pe(s,t)$ & transverse displacement; $\vec r_\pe=\Ord{\epsilon^{1/2}}$ \\
      $r_\pa(s,t)$ & longitudinal displacement; $\vec r_\pa=\Ord{\epsilon}$ \\
      $\varrho(s,t)$ & stored-length density; $\varrho=\vec
      r_\pe'^2/2+\Ord{\epsilon^2}=\Ord{\epsilon}$  \\
      $R$ & end-to-end distance \\
      $R_\pa$ & end-to-end vector projected onto the long.~axis
      \\
      $\simeq$ & equal up to numerical factors of
      order $1$ \\
      $\propto$ & proportional to \\
      $\sim$ & asymptotically equal \\
      $\lp$ &  persistence length \\
      $k_B T$ & thermal energy \\
      $f(s,t)$ & line tension \\
      $\lpe(t)$ &  equilibration scale for transverse bending modes
      \\ 
      $\lpa(t)$ & scale  of tension variations \\
      $\vec \fe$ & external force \\
      $\vec \xi(s,t)$ & thermal force per arc length
    \end{tabular}
  \end{ruledtabular}
\end{table}

\section{Crossover scaling (details)}
\label{sec:cscaling-det}

\subsection{Deterministic relaxation on long times ($t \gg \tf$)}
\label{sec:A}
We have for $\fc=0$
\begin{equation}
  \label{eq:A-step1}
  \begin{split}
    &A-\mint{\frac{dq}{2\pi}}{-\infty}{\infty} \frac 1
    {q^2}\left(1-e^{-2q^2 \phi}\right) = \\
    &\sqrt{\phi}\mint{\frac{dq}{2\pi}}{-\infty}{\infty} \frac 1
    {q^2}\left(e^{-2q^2}-e^{-2q^2(q^2\tau\phi^{-2}+1)}\right) \\
    &\stackrel{\tau\gg1}{\to}0 \;,\qquad \text{if }\phi^{-2}=
    o(\tau^{-1}) \;,
  \end{split}
\end{equation}
and for $\fc=1$
\begin{equation}
  \label{eq:A-step2}
  \begin{split}
    A-\mint{\frac{dq}{2\pi}}{-\infty}{\infty}
    \frac 1 {q^2+1}&=-\mint{\frac{dq}{2\pi}}{-\infty}{\infty}
    \frac 1 {q^2+1} e^{-2q^2(q^2\tau+\phi)}\\
    &\stackrel{\tau\gg1}{\to}0
    \;,\qquad \text{if }\phi^{-2}= o(\tau^{-1}) \;.
  \end{split}
\end{equation}
As indicated, both expressions, Eqs.~(\ref{eq:A-step1},~\ref{eq:A-step2}), go to zero
for large $\tau$ if $\phi^{-2}= o(\tau^{-1})$. The latter, however, follows from the
assumptions stated in the main text, namely that the tension satisfies the scaling
form Eq.~(\ref{eq:scaling-phi}) with the requirement in Eq.~(\ref{eq:alpha-assumption}).
N.b.~it turns out that $\phi=\Ord{\tau}$ for {\pull} and {\lelo} and
$\phi=\Ord{\tau^{4/3}}$ for \epull.

\subsection{Thermal excitation on long times ($t \gg \tf$)}
\label{sec:B}
We want to show that it is justified to calculate (the negative of) the thermally
generated stored length represented by the term $B$, Eq.~(\ref{eq:B}), on long times
\emph{quasi-statically} for the semi-infinite polymer.  To this end, we first
insert the scaling ansatz, Eq.~(\ref{eq:scaling-phi}), for $\phi(\sigma,\tau)$,
\begin{equation}
  \label{eq:B-app-1}
  \begin{split}
    B&=-4 \mint{\frac{dq}{2\pi}}{\Lambda^{-1}}{\infty} q^2 \mint{d\hat
      \tau}{0}{\tau}\\
    &\times e^{-2 q^2 \left[ q^2 (\tau-\hat \tau)+\tau^{\alpha+1}\left(\hat
          \phi(\xi)- (\hat \tau/ \tau)^{\alpha+1}\hat\phi[\xi (\tau/\hat \tau)^{z}
          ]\right)\right] } \;,
  \end{split}
\end{equation}
where we introduced the scaling variable $\xi\equiv
\sigma/\tau^{z}$. Then we substitute $\hat \tau\to x \tau$ and
$q\to q \tau^{-1/4}$,
\begin{equation}
  \label{eq:B-app-2}
  \begin{split}
  B&=-4 \tau^{1/4} \mint{\frac{dq}{2\pi}}{\Lambda^{-1}\tau^{1/4}}{\infty} q^2
  \mint{dx}{0}{1} \\
  &\times e^{-2 q^2 \left[ q^2
        (1-x)+\tau^{\alpha+1/2}\left(\hat \phi(\xi)- x^{\alpha+1} \hat\phi(\xi
          x^{-z})\right)\right] } \;.
  \end{split}
\end{equation}
Note that for $\alpha>-1/2$ (as assumed in Eq.~(\ref{eq:alpha-assumption})) the
factor $\tau^{\alpha+1/2}$ in the exponent diverges in the long-time limit. Hence,
for any given wave number the $x$-integral will be dominated by $x$ close to $1$ for
large enough $\tau\gg1$. This allows us to linearize the exponent in $1-x$ when
performing the $x$-integral for this given wave number. In contrast, for a given time
$\tau$ and $\hat \phi(\xi)=\Ord{1}$, the exponent can be linearized only for large
enough wave numbers, $q\gg q_\star(\tau)\equiv \tau^{-\alpha/2-1/4}$, for which the
factor $q^2 \tau^{\alpha+1/2}\gg1$ in the exponent of Eq.~(\ref{eq:B-app-2}) is much
larger than $1$.

Since the integral over $q$ runs over all $q$-vector, we have also to
care about the small wave numbers, for which the exponent cannot be
linearized in $1-x$. To this end, we split the $q$-integral at a wave
vector $K$ satisfying
\begin{equation}
  \label{eq:K}
  q_\star^{1/3}\gg K\gg q_\star\equiv\tau^{-\alpha/2-1/4} \;,
\end{equation}
which can be found in the limit $\tau\gg1$ under the premise of
Eq.~(\ref{eq:alpha-assumption}), $\alpha>-1/2$.  The first inequality
in Eq.~(\ref{eq:K}) is required for reasons that become clear later
on. For the upper part $B_>$ of the integral we can linearize the
exponent in $1-x$,
\begin{equation}
  \label{eq:B-app-upper}
  \begin{split}
    B_>&\equiv-4 \tau^{1/4} \mint{\frac{dq}{2\pi}}{K}{\infty} q^2
    \mint{dx}{0}{1}\\& \times  e^{-2 q^2 \left[ q^2
        (1-x)+\tau^{\alpha+1/2} \left(\hat \phi(\xi)-x^{\alpha+1} \hat\phi(\xi
          x^{-z})\right)\right] }\\
    &\sim-4 \tau^{1/4} \mint{\frac{dq}{2\pi}}{K}{\infty} q^2
    \mint{dx}{0}{1} \\& \times e^{-2 q^2 (1-x)\left[ q^2
        +\tau^{1/2}\partial_\tau \left(\tau^{\alpha+1}  \hat
          \phi(\sigma/\tau^z)\right)\right] } \\
    &=-2 \tau^{1/4} \mint{\frac{dq}{2\pi}}{K}{\infty}  \frac{1-e^{-2 q^2 \left[ q^2
        +\tau^{1/2}\partial_\tau \left(\tau^{\alpha+1}  \hat
          \phi(\sigma/\tau^z)\right)\right] }}{q^2
        +\tau^{1/2}\partial_\tau \left(\tau^{\alpha+1}  \hat
          \phi(\sigma/\tau^z)\right)} 
  \end{split} 
\end{equation}
where we eliminated the scaling variable $\xi=\sigma/\tau^z$, again.
Using the second inequality in Eq.~(\ref{eq:K}) it is seen that we can
drop the exponential for the ``interesting'' regime
$\hat\phi(\sigma/\tau^z)=\Ord{1}$, where the tension has an
appreciable value. Inserting back
$\phi=\tau^{\alpha+1}\hat\phi(\sigma,\tau)$ we obtain
\begin{equation}
  \label{eq:B-app-upper-2}
  \begin{split}
    B_> &\sim -2 \tau^{1/4} \mint{\frac{dq}{2\pi}}{K}{\infty}
    \frac{1}{q^2
      +\tau^{1/2}\partial_\tau \phi(\sigma,\tau)} \\
    & = \frac{2}{\sqrt{\partial_\tau \phi(\sigma,\tau)}}
    \mint{\frac{dq}{2\pi}}{K/\left(\tau^{1/4}\sqrt{\partial_\tau
      \phi(\sigma,\tau)}\right)}{\infty} \frac{1}{q^2 +1 } \\
    &\sim - \frac{1}{2\sqrt{\partial_\tau \phi(\sigma,\tau)}}\;.
  \end{split}
\end{equation}
To obtain the last asymptotics, we have approximated the lower bound
of the integral by zero. This can be justified by the first inequality
in Eq.~(\ref{eq:K}),
\[\frac{K}{\tau^{1/4}\sqrt{\partial_\tau \phi(\sigma,\tau)}}= K\,
\Ord{\tau^{-\frac 1 4-\frac \alpha 2}=q_\star} \ll q_\star^{4/3}\ll 1 \;.\] 

The remaining lower part $B_<$ of the $q$-integral in
Eq.~(\ref{eq:B-app-2}) is estimated to be small as compared to $B_>$,
\begin{equation}
  \label{eq:B-app-lower}
  \begin{split}
    B_<&\equiv-4 \tau^{1/4} \mint{\frac{dq}{2\pi}}{\Lambda^{-1}t^{1/4}}{K} q^2
    \mint{dx}{0}{1}  \\& \times e^{-2 q^2 \left[ q^2
        (1-x)+\tau^{\alpha+1/2} \left(\hat \phi(\xi)-x^{\alpha+1} \hat\phi(\xi
          x^{-z})\right)\right] }\\
    &< -4 \tau^{1/4} \mint{\frac{dq}{2\pi}}{0}{K} q^2
    \mint{dx}{0}{1} \\
    &= -\frac{4}{3} \tau^{1/4} K^3 /(2 \pi)\\ 
    &\ll -\frac{4}{3} \tau^{-\alpha/2}  /(2 \pi) \\
    &\propto B_> \;,
  \end{split}
\end{equation}
where the first inequality in Eq.~(\ref{eq:K}) has been applied.  Therefore, $B$ is
asymptotically given by Eq.~(\ref{eq:B-app-upper-2}).

\section{Relaxation of a completely stretched polymer}
\label{sec:lelo-app}

In this section, we consider more closely the intermediate asymptotics
\begin{equation}
  \label{eq:homogeneous-lelo}
  f\propto \left(\frac{\hat \zeta L^2}{l_p t}\right)^{2/3} \;, 
\end{equation}
that is approached in the limit
\begin{equation}
  \label{eq:homo-lelo-time-frame}
  \totpa=\frac{\hat \zeta L^2}{l_p {f_<}^{3/2}}\ll t \ll
  t_\star=\frac{\hat \zeta^4 L^8}{l_p^4}\;
\end{equation}
of {\lelo}, which we analyzed in terms of a quasi-static approximation in
Sec.~\ref{sec:tension-relaxed}. The purpose of this section is to justify the
quasi-static assumption in the limit of $f_<\to\infty$ (i.e.~$\tf\to0$) where we
start with a completely stretched polymer and all stored length is generated by the
action of stochastic forces.

To this end, we show that, in the limit $t\to 0$, the change in stored length
$\avg{\Delta\overline \varrho}$ given by Eq.~(\ref{eq:change-stored-length}) for the
force history given by Eq.~(\ref{eq:homogeneous-lelo}) asymptotically approaches the
value one obtains from the quasi-static calculation,
\begin{eqnarray}
  &\avg{\Delta\overline
    \varrho}(t)=\mint{\frac{dq}{\pi\lp}}{0}{\infty}
  \left\{\frac{1}{q^2+f_<}\left(e^{-2q^2[q^2 t+ F(t)]}-1\right)
  \right.& \nonumber\\
  &\left.+2 q^2\mint{d\tilde t}{0}{t}e^{-2q^2\left[q^2(t-\tilde
        t)+F(t)-F(\tilde t)\right]}\right\}&\nonumber\\
  &\stackrel{f_<\to\infty}{=}\mint{\frac{dq}{\pi\lp}}{0}{\infty} 2
  q^2\mint{d\tilde t}{0}{t}e^{-2q^2\left[q^2(t-\tilde
      t)+F(t)-F(\tilde t)\right]} & \label{eq:to-show-1} \\
  &\stackrel{!}{\sim} \mint{\frac{dq}{\pi\lp}}{0}{\infty}
  \frac{1}{q^2+f(t)}=\frac{1}{2l_p}\left[f(t)\right]^{-1/2}\;,\qquad\text{for } t
  \to 0\;.&\nonumber
\end{eqnarray}
This will comprise an a posteriori justification of the quasi-static
assumption that entered in the derivation of the right-hand-side of
Eq.~(\ref{eq:degennes-eq-2}).

The argument closely follows App.~\ref{sec:B}. Inserting the force history
\begin{equation}
  \label{eq:F-1/3}
  F(t)=\mint{d\hat t}{0}{t}f(\hat t)= C t^{\alpha+1}
\end{equation}
with
\begin{equation}
  \label{eq:prefactor}
  \alpha=-2/3\stackrel{!}{<}-1/2 \qquad \text{and}\qquad C\approx \left(\frac{\hat \zeta L^2}{l_p}\right)^{2/3}
\end{equation}
into Eq.~(\ref{eq:to-show-1}) and changing variables $\hat t \to x t$
and $q\to q t^{-1/4}$ yields
\begin{eqnarray}
  \label{eq:1/3-noise-rescaled}
  &&\avg{\Delta\overline
      \varrho}(t)=2t^{1/4}\mint{\frac{dq}{\pi\lp}}{\Lambda^{-1}t^{1/4}}{\infty} 
    q^2  \nonumber \\
    &&\qquad\times \mint{dx}{0}{1}e^{-2q^2\left[q^2(1-x)+C t^{\alpha+1/2}\left(1-x^{\alpha+1}\right)\right]}
\end{eqnarray}
As in Sec.~\ref{sec:B} an approximation to the integral can be found in
the limit
\begin{equation}
  \label{eq:central-limit}
  C t^{\alpha+1/2}\gg1 \;.
\end{equation}
We split the $q$-integral at $K$ satisfying
\begin{equation}
  \label{eq:K2}
  q_\star^{1/3}\gg K\gg q_\star\equiv (C t^{\alpha+1/2})^{-1/2} \;,
\end{equation}
which can be found in the limit $t\ll 1$ ($\Rightarrow
q_\star\ll1$) because $\alpha<-1/2$. The upper part of
the integral
\begin{equation}
  \label{eq:rho>}
  \avg{\Delta\overline
      \varrho^>}(t)\equiv \left(\dots\right)\mint{dq}{K}{\infty}\left(\dots\right)
\end{equation}
is dominated by values of $x$ close to $1$ and we can linearize the
exponent in $1-x$,
\begin{equation}
  \label{eq:rho>-2}
  \begin{split}
    \avg{\Delta\overline{
        \varrho}^>}(t)&=2t^{1/4}\mint{\frac{dq}{\pi}}{K}{\infty}q^2\mint{dx}{0}{1}
    \\&\times e^{-2q^2 (1-x) \left[q^2+(\alpha+1)C t^{\alpha+1/2} \right]} \\
    &= \tau^{1/4}\mint{\frac{dq}{\pi}}{K}{\infty} \frac{1-e^{-2q^2 (1-x)
        \left[q^2+(\alpha+1)C t^{\alpha+1/2}
        \right]}} {q^2+(\alpha+1)C t^{\alpha+1/2}} \\
    &\sim \tau^{1/4}\mint{\frac{dq}{\pi}}{K}{\infty}
    \frac{1}{q^2+(\alpha+1)C t^{\alpha+1/2}} \\
    &=\frac{1}{l_p\sqrt{(\alpha+1)C t^\alpha}}\mint{\frac{dq}{1
        \pi}}{K/\sqrt{(\alpha+1)C t^{\alpha+1/2}}}{\infty}\frac{1}{q^2+1} \\
    &\sim \frac{1}{2 l_p \sqrt{f(t)}}
\end{split}
\end{equation}
where the asymptotics follow from both inequalities in
Eqs.~(\ref{eq:central-limit},~\ref{eq:K2}). The lower part $\avg{\Delta\overline{
    \varrho}^<}$ of the integral is estimated to be subdominant as compared to
$\avg{\Delta\overline{ \varrho}^>}$,
\begin{equation}
  \label{eq:rho<}
  \begin{split}
    \avg{\Delta\overline{
      \varrho}^<}(t)&=(\dots)\mint{dq}{0}{K}(\dots) \\
    &<\frac{2 t^{1/4}}{l_p}\mint{\frac{dq}{\pi}}{0}{K} q^2
    \mint{dx}{0}{1} \\
    &=\frac{t^{1/4}K^3}{3\pi l_p}\\
    &\ll\frac{C^{-1/2}t^{-\alpha/2}}{3\pi l_p} \\
    &\propto \avg{\Delta\overline{
      \varrho}^>}(t) \;,
  \end{split}
\end{equation}
where the first inequality in Eq.~(\ref{eq:K2}) has been applied. Therefore
$\avg{\Delta\overline{ \varrho}}(t)$ is asymptotically given by
Eq.~(\ref{eq:rho>-2}).

Finally, we want to emphasize the central condition for the validity
of the quasi-static approximation,
\begin{equation}
  \label{eq:central-limit-2}
  C t^{\alpha+1/2}\approx\left(\frac{\hat \zeta L^2}{l_p}\right)^{2/3}t^{-1/6}\gg 1\;,
\end{equation}
or $t\ll t_\star$ with $t_\star=(L^2/\lp)^4$ as given by
Eq.~(\ref{eq:val-cond-lelo-2}).

\section{Defining $t_\star$}
\label{sec:tstar}

As anticipated Part~I, there is a problem-specific time $t_\star$ limiting the
short-time validity of OPT.  Physically, the crossover at $t_\star$ can be understood
as follows. For $t\gg t_\star$ the ``speed'' of the structural relaxation is
determined solely by the relaxation times of the bending modes, which are related to
the transverse friction while the longitudinal friction is irrelevant. In contrast,
for $t\ll t_\star$ the longitudinal friction substantially limits the speed of the
relaxation. This suggests to estimate the time $t_\star$ as follows.  From the
continuity equation, Eq.~(\ref{eq:cg-eom}), derived via the multiple-scale
perturbation theory (MSPT), we can estimate the order of magnitude of the correction
$\delta f(s,t)=f(s,t)-f^{\text{OPT}}$ to the flat tension profile,
Eq.~(\ref{eq:OPT-tension}), by
\begin{equation}
  \label{eq:f-correction}
  \delta f \approx \hat \zeta \partial_t \avg{\Delta \overline{\varrho}}\left(
    f^{\text{OPT}},t  \right) L^2 \;.
\end{equation}
OPT can only be applicable if the correction $\delta f$ has negligible
effect on the evolution of the stored length,
\begin{equation}
  \label{eq:OPT-validity}
  \abs{1-\frac{\avg{\Delta\overline{\varrho}}\left(
    f^{\text{OPT}}+\delta f,t  \right)}{\avg{\Delta\overline{\varrho}}\left(
    f^{\text{OPT}},t  \right)}}\ll 1 \;,\qquad \text{for } t\gg t_\star\;.
\end{equation}
The time for which the left hand side of Eq.~(\ref{eq:OPT-validity}) becomes of order
may thus be identified with the time $t_\star$ before which OPT is not valid.

\bibliographystyle{apsrev}

\bibliography{bibis/tdpre06,bibis/elastic-rod-dynamics,bibis/journals,bibis/sfpdynamics-tcited,bibis/unpub,bibis/actin-viscoelastic,bibis/semiflexibleA04,bibis/sf,bibis/klaussf,bibis/mysf,bibis/sfnet,bibis/unpub,bibis/mysfnet}

\end{document}